\documentclass[aps,prb,twocolumn,superscriptaddress]{revtex4-2}

\usepackage{graphicx}
\usepackage{amsmath}
\usepackage{amsfonts}
\usepackage{color}
\usepackage{listings}
\usepackage{xcolor} 
\usepackage{hyperref}
\newcommand{\mm}[1]{\textcolor{red}{#1}}

\usepackage[normalem]{ulem} 

\begin{document}

\title{Finite-temperature properties of the Frenkel-Kontorova model:\\
Relation to tribological systems and fluid rheology}

\author{Shubham Agarwal}
\affiliation{Dept. of Materials Science and Engineering, Campus C6 3, Saarland University, 66123 Saarbrücken, Germany}
\author{Martin H. Müser}
\affiliation{Dept. of Materials Science and Engineering, Campus C6 3, Saarland University, 66123 Saarbrücken, Germany}
\email{martin.mueser@mx.uni-saarland.de}

\date{\today}

\begin{abstract}

The Frenkel-Kontorova model is a simple yet generic framework for the description of tribological phenomena and processes, including dry solid friction and the motion of adsorbed layers.
As revealed in this work, it also reproduces qualitatively various features of complex liquids, such as, power-law sub-diffusion between the ballistic and the diffusive regimes as well as a cross-over from a non-Arrhenius to an Arrhenius dependence of the diffusion coefficient near the temperature, where the specific heat assumes its maximum. 
The study of these and related thermal and kinetic properties highlights several misconceptions prevalent in the literature.
Most notably, 
%
shear thinning with a shear-thinning exponent close to zero can be the natural consequence from enforced basin hopping:
the energy drops caused by shear-induced instabilities dictate the friction-velocity dependence at medium shear rates rather than the way how shear forces reduce the free energy barriers for directed motion. 
Thus, even if the rheology is described by semi-empirical theories such as the Eyring model, any agreement with experimental data, whether past, present, or future,  may be purely coincidental.
\end{abstract}

\maketitle

\section{Introduction}

The Frenkel Kontorova (FK) model, which consists of a linearly elastic chain that is placed into an external, periodic potential and driven by an external force, was originally introduced in 1938 by Yakov Frenkel and Tatiana Kontorova to study dislocation dynamics in crystals~\cite{Frenkel1938}.
Since then, it has found many applications ranging from charge-density waves to solid friction~\cite{Flora1996AP,Braun1997PRE,Vanossi2007JP,Strunz1998PRE,Strunz1998bPRE}.
However, the FK model may require substantial generalization to be quantitative such as incorporating the non-local elasticity of surface modes to account for half-space elasticity~\cite{Peierls1940PPS,Nabarro1947PPS}.
It nevertheless allows trends to be rationalized quite generally even in its original form, because it features a competition between (elastic) restoring and (periodic) external forces.
As such it perfectly addresses the microscopic mechanisms that Coulomb~\cite{Coulomb85} had already envisioned to cause friction: the coherence that molecules adopt at solid interfaces due to their proximity, which must be overcome to initiate sliding as well as the vibrations caused when asperities of opposing surfaces move past each other. 

In a tribological context~\cite{Muser2003ACP}, the FK model has been primarily studied in a surprisingly narrow parameter range, i.e., at high speeds near the velocity of sound~\cite{Strunz1998PRE,Strunz1998bPRE}, or near the transition from Stokes-like to Coulomb-like friction~\cite{Vanossi2007JP, Benassi2015SR, vandenEnde2012}, which relates, for real systems, to situations where stiff solids with passivated surfaces slide past each other~\cite{Muser2004EPL}.
While successful attempts were made to adjust the parameters of primarily two-dimensional FK models to experiments of model systems like adsorbed, colloidal~\cite{Brazda2018PRX} or atomic monolayers~\cite{Norell2016PRE},
relatively, little appears to be known when thermal effects matter at small sliding velocities and when the stiffness of the springs connecting two adjacent beads is of similar size as the maximum curvature of the substrate potential.
Viewed from a different angle, the FK model has not yet been explored in the context of fluid rheology, most notably to investigate whether generic explanations can be identified that account for the non-Arrhenius dependence of viscosity~\cite{Angell1991JNCS} or the reasons why shear thinning sometimes obeys the Eyring model~\cite{Eyring1936JCP,Spikes2015TL} but more generally the phenomenological Carreau-Yasuda equation~\cite{Carreau1972TSR,Yasuda1981RA}.
The aim of this article is to fill this gap.

While relating the FK model to fluid rheology may appear far-fetched at first sight, laminar flow is dissipative also due to the competition between the restoring forces within a lamella and those between them, as explored, for example, in a study of an athermal FK model sandwiched between moving incommensurate walls~\cite{Vanossi2008TI}.
Moreover, the local structure in liquids can be stable over many vibrational periods, even in low-viscosity liquids like water at ambient conditions~\cite{Rapaport1983MP}.
Its mass transport is intimately linked to the formation and propagation of coordination defects breaking the tetrahedral network formed by water molecules~\cite{Kuhne2007PRL}. 
It could be argued that the discommensurations in the FK model, which are local deviations from having one atom per external potential minimum, are caricatures of such defects. 
Although improving our understanding of the \emph{thermal} properties of the FK model -- which is traditionally studied at zero temperature~\cite{Aubry1983PDNL,Aubry1983PDNL-2} -- is interesting in its own right, the presented study is motivated by our desire to rationalize typical liquid rheology in simple terms through a microscopic model.
Commonly pursued approaches fail in this regard, because they are either intransparent and formalism laden, such as mode coupling theory~\cite{Bengtzelius1984JPC}, and/or clearly insufficient in that their predictions fail to describe the shear thinning of many liquids.  
Moreover, pseudo-microscopic approaches, such as Eyring theory when applied to the FK model, intrinsically fail to distinguish between situations, in which the chain is driven either by distributing the external force equally among the beads in the chain or by applying the entire bead merely to one front bead.
Yet, the effective damping or viscosity -- e.g., the ratio of shear force and velocity -- can depend crucially on such details.

The static ground-state properties of the generic, classical FK model
is usually discussed in terms of its two most important dimensionless numbers~\cite{Aubry1983PDNL,Aubry1983PDNL-2,Braun1997PRE}: the misfit between the substrate period and the intrinsic length of the spring connecting two adjacent beads, and the ratio of spring stiffness to the maximum substrate potential curvature.
However, quite a few additional parameters affect \emph{qualitatively} how sliding velocity depends on additional (dimensionless) parameters defining the model. 
This includes, but is not limited to, aspects already alluded to in the discussion so far -- e.g., local vs. non-local elasticity, temperature, the way how the external force is imposed, and inertial versus damped dynamics -- but also the way how damping is applied, i.e., relative to the substrate or in a momentum-conserving fashion between adjacent beads, and periodic boundary conditions versus open-chain dynamics, etc.
Thus, there is a plethora of limiting cases that can be studied in principle.
An important question then is: which parameter choices reflect specific systems most realistically?
Studying, as occasionally happens, resonance phenomena at sliding velocities close to the speed of sound within the chain will scarcely address experimentally relevant observations in a practical tribological context. 
Thus, in this work, we attempt to scrutinize how different model choices can affect the friction of a Frenkel-Kontorova chain and interpret the results under the premise that the model was constructed to mimic the rheology of a bulk liquid or potentially of a boundary lubricant at high pressure. 

We also wish to understand better how the FK model relates to its mean-field limit, the Prandtl model~\cite{Prandtl1928ZAMM,Popov2012ZAMM}.
The latter assumes elastic coupling of atoms to their ideal lattice sites, resulting in interactions that are too long-ranged for semi-infinite solids, while the FK model represents the opposite limit with interactions that are too short-ranged.
On one hand, the Prandtl model closely obeys power-law shear thinning as described by the Carreau–Yasuda equation~\cite{Muser2020L}, capturing both temperature and rate dependence of the viscosity of simple alkanes, which suggests that some elementary instability underlies shear thinning in these systems~\cite{Gao2023TL}.
On the other hand, evidence for Eyring-type behavior continues to emerge~\cite{Manzi2014TL,Manzi2021TL}.
However, the Prandtl model alone is not satisfactory, as it neglects collective dynamics and predicts that only the static friction, not the steady-state kinetic friction, depends on the degree of incommensurability~\cite{Muser2003ACP}.
In real materials, plasticity and flow arise not merely from individual atomic motions but from cooperative rearrangements.
This collective behavior—central to phenomena such as dislocation motion and defect nucleation—is inherently absent in a single-degree-of-freedom model like Prandtl’s.
To bridge this gap, we turn to the FK model, which incorporates many-body interactions and spatial correlations, enabling the study of collective effects in a minimal framework.

%


\section{Background, model and methods}

\subsection{Historical note}
Physics 101 text books but also books and reviews in tribology often claim that Coulomb found kinetic friction to be independent of sliding velocity. 
This is only partly true, since he made such a statement mainly for contact between metals. 
For contacts formed by wood and metals, he found
\textit{
le frottement croît très-sensiblement à mesure que l'on augmente les vîtesses; en sorte que le frottement croît à peu près suivant une progression arithmétique, lorsque les vîtesses croissent suivant une progression géométrique
}~\cite{Coulomb85}.
The translation to English would be roughly
\textit{
Friction increases very noticeably as speed increases; in a way that friction increases approximately in an arithmetic progression when speed increases in a geometric progression.
}
In other words, if friction increases by a constant amount  with each $\Delta F$ increment in the force, $F_{n} = F_0 + n \Delta F$, then velocity scales by a constant factor in each increment so that $v_{n} = c^n v_0$.
Solving for $n$ in both cases, equating the results, and dropping the index $n$ yields
\begin{equation}
F = F_0 + \frac{\Delta F}{\ln c} \ln\frac{v}{v_0}.
\end{equation}

Since the inverse hyperbolic sine quickly converges to the natural logarithm at large arguments, Coulomb's observation is in line with Eyring theory, albeit his explanation of the phenomenon not necessarily so. 
He argued
\textit{Plus la vitesse sera grande, plus il faudra plier de fois la fibre... à mesure que la vitesse augmentera, parce qu'en passant d'une sommité à l'autre, les fibres n'ont pas le temps de se redresser en entier.}
This translates roughly to
\textit{
The greater the speed, the more strongly the [wood] fiber[s] must be bent ... at large speeds, the fibers do not have enough time to fully straighten when passing from one peak to the next.}
The processes that Coulomb alluded to here are entirely mechanical not thermal.
In todays jargon, they could be related to the (potentially non-linear) visco-elastic response of the fibers.
The mechanistic picture provided by Coulomb can be easily generalized to other situations, e.g., by replacing the wood fibers with elastic strings in the FK model, or, taking it one step further, with discommensurations or even dislocations.
This opens the question to what degree shear thinning is a mechanical effect, as envisioned by Coulomb, or thermodynamic as argued by Eyring. 

\subsection{Model}

\subsubsection{Default model}

The default Frenkel-Kontorova model studied in this work consists of a one-dimensional, linearly harmonic bead-spring chain with nearest-neighbor interactions, which is placed into a single-sinusoidal potential of the form $V(x) = V_0 \cos(qx)$.
Here, $V_0$ is of unit energy and $q = 2\pi/b$ is the wavenumber of the potential and $b$ the period. 
Damping forces linear in velocity act on individual beads, leading to the equation of motion for a bead within the chain 
\begin{equation}
\label{eq:fk_model}
m \ddot{u}_n + \frac{m}{\tau} \dot u_n = k ( x_{n+1} + x_{n-1} - 2 u_x) +
q V_0 \sin(qx_n)+ \Gamma_n(t).
\end{equation}
Here, $m$ is the mass of a bead, $\tau$ is a relaxation time, and $k$ is the sping stiffness connecting two adjacent beads, while $n = 0,\dots,P-1$ enumerates the beads. 
$\Gamma_n(t)$ is a thermal random force with a mean value of zero and second moments of
\begin{equation}
\left\langle \Gamma_m(t) \Gamma_n(t') \right\rangle =  2 k_B T m
\delta_{mn} \delta(t-t') / \tau,
\end{equation}
where $k_BT$ is the thermal energy. 

The two terminating beads are treated depending on the boundary condition.
End beads of an open chain are only coupled to one neighbor.
%
In the case of periodic boundary conditions (pbc), the first and last bead subject a force of 
\begin{equation}
F_{0,P-1} = \mp k (x_{0} - x_{P-1} + L - a)
\end{equation}
onto each other, where $a = L/P$ would be the preferred spacing beads in a free chain, just like in an open chain.

The FK model was introduced as a crude model for the motion of dislocations, $\vert b/a - 1\vert \ll 1$, or of grain boundaries with severe less strict restrictions for the $b/a$ ratio. 
%
%
More recently, it was used to model frictional interfaces between atomically smooth, uncontaminated surfaces~\cite{Strunz1998PRE,Strunz1998bPRE,Braun2001PRE, Vanossi2007JP}.
In each of these applications, a quantitative approach would need to properly reflect non-local elasticity, which, however, has barely been pursued so far~\cite{Hirano1990PRB}.
Non-local elasticity does not matter in systems lacking a half space in the direction normal to the interface, such as multiwalled carbon nanotubes or adsorbed layers, though the latter require a second in-plane dimension to be introduced.
While the FK model has scarcely been discussed in the context of polymer dynamics, one could argue that it also represents some of the aspects qualtiatively, albeit potentially in a more coarse-grained spirit, e.g., for the motion of a polymer past a surface or along its tube of constraints in the reptation model. 
Depending on context, different parameterizations of the FK model might be appropriate.
The wealth of possibilities is extremely large, simply because the model has quite a few dimensionless constants, despite its simplicity.

In the FK model variant introduced in Eq.~\eqref{eq:fk_model}, three parameters can be set unity.
We chose the bead inertia $m$, the corrugation barrier $V_0$, and the equilibrium spacing between beads in the chain $a$, which define our unit system. 
In the following, we will indicate variables that are expressed in this unit system with a tilde, e.g., $\tilde{b} = b/a$.
The other critical non-dimensional parameter of the FK model is the dimensionless number $\tilde{k} \equiv ka^2/V_0$.
However, the spring stiffness is sometimes more meaningfully undimensionalized through $k^* \equiv k/V''_\textrm{max}$, where $V''_\text{max} = q^2 V_0$ is the maximum curvature of the substrate potential. 
Throughout this work $k^*/\tilde{k} \approx 0.02882 \approx 1/35$.

It is well explored how the precise values of $\tilde{b}$ and $\tilde{k}$ affect the ability of the chain to interlock with substrate.
The effect of the remaining dimensionless parameters and model choices has been explored substantially less.
In most studies, thermal fluctuations are ignored, however, the reduced temperature $\tilde{T} = k_B T / V_0$ may well be finite and depending on its precise value, qualitatively different properties of the chain can ensue.
The damping $m/\tau$ can make the chain be strongly overdamped or underdamped or (near) critically damped. 
The damping itself is not necessarily relative to the substrate but could also be envisioned to happen within the springs. 
Similarly important, the number of beads in the chain can take a small (order unity), medium or large value.
Last, but not least, periodic boundary conditions can be on or off.
This allows a plethora of FK model classes to be constructed whose properties can differ from each other qualitatively, e.g., Arrhenius vs. non-Arrhenius viscosity, Eyring vs. non-Eyring rheology, presence or absence of shear localization at high velocities, i.e., shear thinning exponents above or below unity, and the existance or absence of resonance friction at intermediate sliding velocities, to list the most important properties of FK chains discussed in this work. 

In the majority of cases, we use $m = 1$ and $\tau = 1$, in particular when running simulations at finite temperature or when using a constant lateral force. 
However, when studying athermal sliding at small velocities, we usually use a smaller relaxation time of $\tau = 0.2$ to suppress oscillations in the dynamics, which occur after an instability in an underdamped system was triggered. 
These oscillations are visual clutter reducing the readability of the figures.

Further deviations from default settings will be detailed either in Sect.~\ref{sec:unconventional} or on place in the results section. 
At the same time, it is necessary to restrict the variation of parameters.
This is why we consider a fixed number of atoms per chain and a fixed $a/b$ ratio, specifically $P = 16$ and  $a/b = 15/16$. 
This number of $P = 16$ is large compared to one, but small enough so that the elastic object could crudely represent either hexadecane or a nano-crystal. 
At the same time, $a/b$ could be interpreted as an extended small-angle grain boundary with an angle of about $360^\circ/(2 \pi \cdot 16) \lesssim 5^\circ$. 
While attempting to establish rough relations between parameterizations and real systems, the original motivation of this work is to learn generic reasons for why Arrhenius and Eyring model appear to be valid in some situations but not in others. 

The equations of motion were solved using an in-house written Python code, which is available on Github \cite{Muser2025Github}.
For the majority of simulations, i.e., when default damping applied, the Grønbech-Jensen thermostat~\cite{GronbechJensen2019MP} is used, while a colored and a momentum-conserving thermostat, described elsewhere~\cite{Agarwal2025arxiv}, are used for damping schemes, which are described next. 

\subsubsection{Alternative damping schemes}  
\label{sec:unconventional}

In addition to the default model defined in Eq.~\eqref{eq:fk_model}, we also consider two alternative approaches for dissipating energy, which deviate from the standard implementation of instantaneous damping relative to the substrate.

In the first approach, damping and thermalization are implemented via Maxwell elements. This corresponds formally to a non-Markovian damping kernel, where both friction and random forces act with a finite delay rather than instantaneously. For further details on this class of thermostats, we refer to a recent study~\cite{Agarwal2025arxiv}.

In the second approach, energy is dissipated through damping of the relative motion between adjacent beads in the chain. The associated random force compensates this damping, preserving thermal equilibrium. The resulting dynamics is similar to dissipative particle dynamics~\cite{Espanol1995EPL,Soddemann2003PRE,Vattulainen2002JCP} but implemented through a recently proposed momentum-conserving Langevin thermostat~\cite{Agarwal2025arxiv}, which transitions seamlessly from Langevin to Brownian dynamics, while allowing the time step to increase when damping increases at fixed mass. 

Physically, the Maxwell-element model captures a viscoelastic response of the substrate, possibly originating from unresolved internal degrees of freedom. In contrast, the bead-bead damping model implies that energy dissipation occurs within the chain itself, representing a coarse-grained description of a soft, dissipative medium sliding over a rigid substrate.

\section{Results}

\subsection{Thermal FK model in the extreme small-coupling limit}

In this work, we refer to small coupling when the \emph{elastic} coupling between adjacant beads is small, that is, when the spring stiffness $k$ is small compared to the maximum curvature of the potential $\kappa = q^2 V_0$. 
In the extreme limit $k^* \equiv k/\kappa \to 0^+$, the motion of particles is primarily dictated by the substrate potential.
However, the thermodynamics of any positive $k$ still differs qualitatively from $k = 0$, as the configurational specific heat will assume the value of $c_\textrm{c}(T\to\infty) = (1-1/N) k_B/2$ at large $T$ while that of uncoupled beads would vanish with increasing temperature.
In the latter case, $c_\text{c}$ can be calculated analytically from the (configurational) partition function $z_\textrm{c}(\beta)$ 
\begin{eqnarray}
z_\textrm{c}(\beta) & \propto  &
\int_{0}^{2\pi/q} \mathrm{d}x e^{-\beta V_0 \cos(qx)} \\
& \propto & I_0(\beta V_0).
\end{eqnarray}
Here $\beta = 1/k_B T$ is the inverse thermal energy, while $I_n(x)$ is the modified Bessel function of the first kind of order $n$.
We have integrated over one period, as the integral over more periods only leads to an irrelevant prefector, however, large it may be. 
Taking the negative  derivative of $\ln z_c(\beta)$ with respect to $\beta$ gives the (configurational) internal energy per atom
\begin{equation}
u_{\text{c}}(\beta) = -V_0 \frac{I_1(\beta V_0)}{I_0(\beta V_0)}.
\end{equation}
The derivative of $u_{\text{c}}(\beta)$ with respect to $T$ then yields the specific heat 
\begin{equation}
c_{\text{c}}(T) = \frac{V_0^2}{k_B T^2} \,
\frac{I_0(x)(I_0(x) + I_2(x))/2-I_1^2(x)}{I^2_0(x)} . 
\end{equation}
This result has certainly be identified prior to this work, even if suitable references could not be indentified.

\begin{figure}[htbp]
\includegraphics[width=0.9\columnwidth]{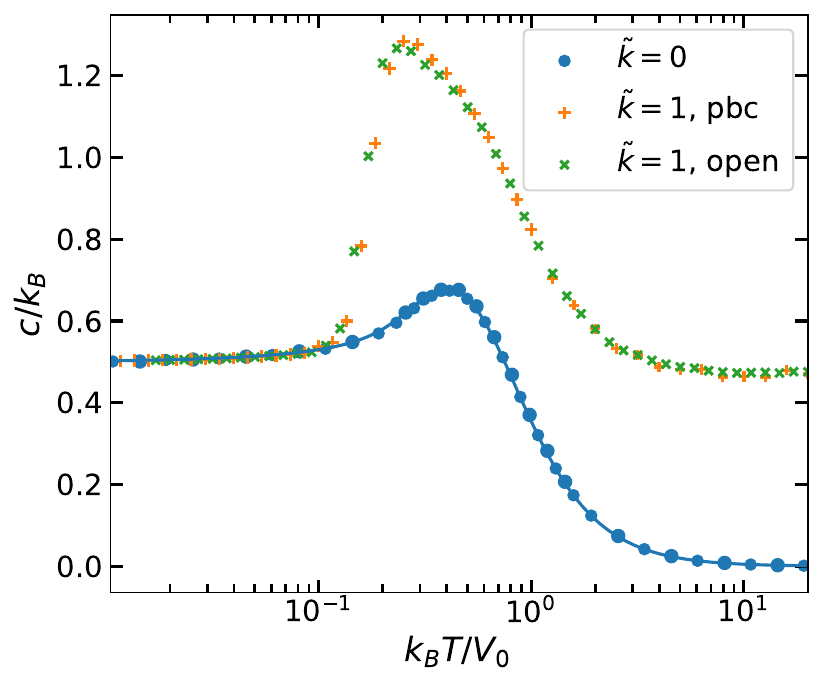}
\caption{\label{fig:cp_free}
Configurational specific heat $c_\text{c}$ as a function of reduced temperature $k_BT/V_0$ for free atoms (blue circles) and  weak springs, $k = V_0/a^2$, which can be open (green crosses) or subjected to periodic boundary conditions (pbc, orange plus symbols). 
}
\end{figure}

Fig.~\ref{fig:cp_free} reveals consistency between the analytical and numerical results. 
%
Fig.~\ref{fig:cp_free} also contains data for weak (elastic) coupling, i.e., $k = V_0/a^2$, which translates to a dimensionless coupling constant of $k^* = 2.882\cdot 10^{-2}$.
As the zero-coupling limit, the weak-coupling system obeys the rule of Dulong-Petit at small temperature and the specific heat assumes its maximum again near $\tilde{T}\ = 0.4$.
However, the peak extends to both larger and smaller temperatures than before.
This is because chains with small $\tilde{k}$ are multistable with many inequivalent minima energy minima.
In contrast, all energy minima are equivalent for uncoupled beads in the chain.
Since different energy minima can be separated by large energy barriers, proper sampling at small temperature turns out difficult.
This explains the relatively large scatter in the specific heat at small temperature, which persisted even when using more than 50 million time steps and averaging over 32 independent samples.
%
%

\begin{figure}[htbp]
\includegraphics[width=0.9\columnwidth]{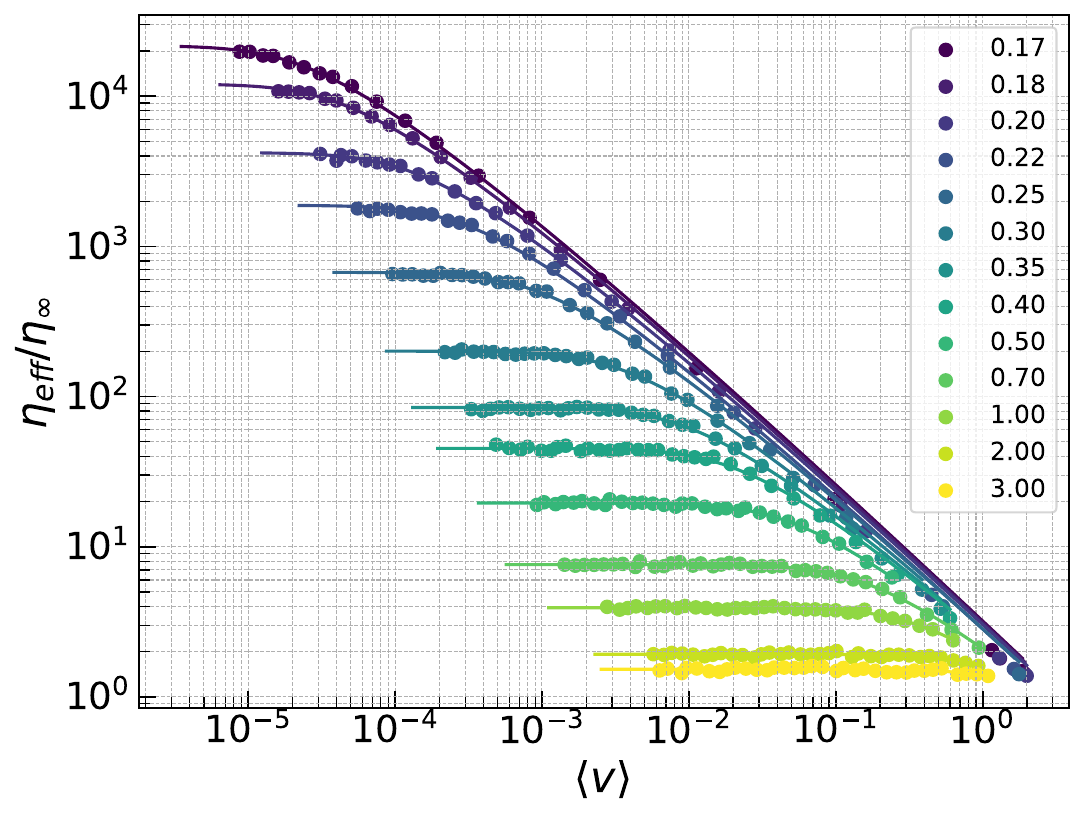}
\caption{\label{fig:eff_visc_free}
Effective viscosity of free atoms in a corrugated potential.
Lines reflect fits to the Eyring equation. 
}
\end{figure}

The rheological response of free and weakly bonded atoms were determined with simulations in which a constant temperature $T$ and constant external force $F$ were applied and the mean slid distance $d$ measured.
This was done by setting up $N = 32$ chains in parallel and by running the simulations long enough so that a target standard deviation in the measured slid distance of 2.5\% was met, once the chains had slid a mean distance exceeding $2 b$. 
The net damping, which we call effective viscosity $\eta_\text{eff}$, is then computed as $\eta_\text{eff} = F/\langle v \rangle$, where $\langle v \rangle$ is the ratio of $d$ and the simulation time $t_\textrm{sim}$ spent to reach the target accuracy. 
In the following simulation, the external force was reduced by a factor equal or close to $\sqrt[10]{0.1}$.
The procedure is repeated until the number of time steps needed to reach convergence exceeds a given threshold value, which was typically of the order of a few hundred million time steps. 
Numerical results are presented in Fig.~\ref{fig:eff_visc_free}
along with fits to the Eyring model
\begin{equation}
\label{eq:eyring}
\eta_\text{eff} = \eta_0 \frac{v_0}{v}
\text{asinh}\left( \frac{v}{v_0} \right),
\end{equation}
where $\eta_0$ is the equilibrium viscosity and $v_0$ a velocity near which deviations from equilibrium viscosities become substantial. 

As is the case for some liquids under certain conditions, in particular at high pressure and small temperature, Eyring provides a quite satisfactory description over a few decades in velocity, which, would usually be a few decades in shear rate. 
The given case constitutes an ideal reference for Eyring theory, because both the force acting on the degree of freedom to pass the barrier and the barrier itself are constants in time. 

The validity of Eyring theory for the reference case is well established and text-book material~\cite{Risken1989book}.
In brief, at very low temperature, atoms are located near the potential energy minimum most of the time. 
The barrier in forward and backward direction read $\Delta E_{\pm} = 2 V_0 \mp F b/2$ so that a net flow velocity (roughly) proportional to $e^{-2 V_0/k_B T}\sinh(Fb/2k_BT)$ follows, where the proportionality coefficient, $v'_0(T)$, is supposed to depend only weakly, e.g., algebraically but not exponentially,  on temperature
\begin{equation}
\label{eq:FK_thermal}
v(F,T) = v_0'(T) e^{-2 V_0/k_B T}\sinh(Fb/2k_BT).
\end{equation}

The main trouble maker in Eq.~\eqref{eq:FK_thermal} or more generally speaking in transition-state theory~\cite{Kramers1940P,Hanggi1990PRM} is the function $v_0'(T)$.
In the high-temperature limit, it must approach $v_0' \approx b / (2 k_B T \gamma)$ in our case so that  $\eta_\infty$ corresponds to the explicitly imposed damping. 
At low temperature, $v_0'(T)$ assumes a slightly greater value, which depends on the details of the model, e.g., the precise shape of the potential, to what degree the motion is under- or overdamped but also the collectivity of the barrier-crossing process. 
Various established approaches exist to correct the leading-order approximation contained in Eq.~\eqref{eq:FK_thermal}~\cite{Hanggi1990PRM,Risken1989book} in different asymptotic limits, such as overdamped motion in a single sinusoidal potential~\cite{Ambegaokar1969PRL}.

At medium to high velocity, the Eyring model is no longer highly accurate, at least when the damping is large.
This is mostly because of inertia, which allows atoms to use a fraction of their kinetic energy released during the last barrier crossing to overcome the next barrier. 
This reduces the effective viscosity.
At very high velocity, the explicit damping counteracts this effect, which, however, can be accounted for by adding $\eta_\infty$ to the Eyring viscosity. 

\begin{figure}[htbp]
\includegraphics[width=0.9\columnwidth]{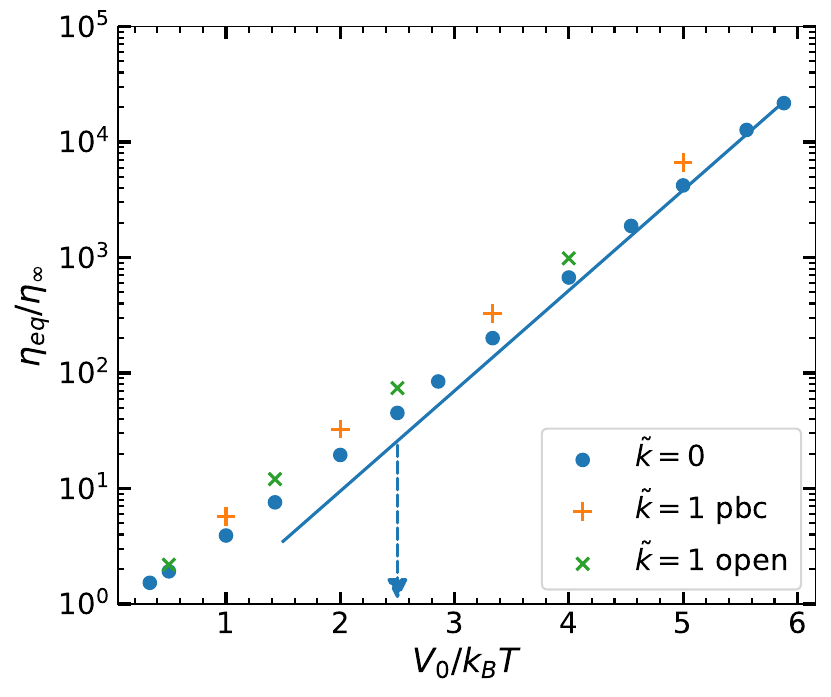}
\caption{\label{fig:eq_visc_free}
Equilibrium viscosity as a function of temperature for free atoms (blue circles).
The slope of the solid line reflects the activation energy of $\Delta E = 2~V_0$.
%
The arrow indicates the temperature, at which the specific heat assumes its maximum. 
Data for weakly coupled chains is included, specifically for $k a_0^2 / V_0 = 1$,
one time with open (green crosses) and one time for closed (orange plus symbols). 
}
\end{figure}

We next investigate the equilibrium viscosity as a function of inverse temperature in Fig.~\ref{fig:eq_visc_free}.
The data show clear Arrhenius behavior at low temperature and minor deviations from it  near or above temperatures at which the specific heat assumes its maximum. 
The departure of Arrhenius behavior at large temperatures is contained in analytical solutions~\cite{Ambegaokar1969PRL} and might be argued to occur, because atoms no longer typically find themselves near the potential minima when temperature is high, which can reduce $\Delta E$ at large $T$. 
%
%
%
%

Introducing a weak coupling of $k = 0.1 V_0/a^2$  (dimensionless coupling of ${k}^* \approx 2.882\cdot 10^{-3}$) somewhat increases the equilibrium viscosity $\eta_0$ but leaves the apparent activation energy $\Delta E_\text{app} = - \partial \ln \eta_0 / \partial \beta$ almost unaltered. 
Thus, weak coupling has only minor effects on both equilibrium and effective viscosity, the latter being not shown, although it alters the specific heat substantially. 

%
%
%

%
%
%

\subsection{Periodic FK chain at medium elastic coupling}

\begin{figure}[htbp]
\includegraphics[width=0.9\columnwidth]{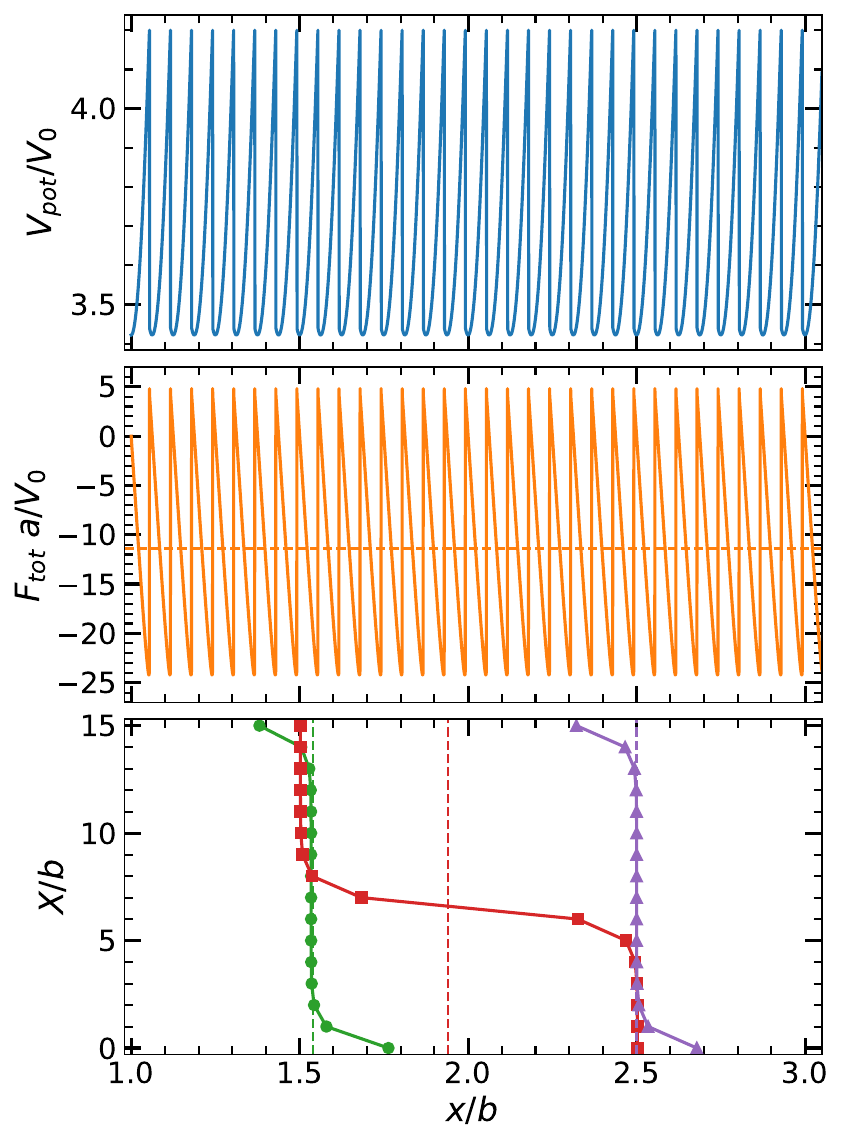}
\caption{\label{fig:barr10pbc}
Athermal dynamics of a $\tilde{k} = 10$ chain subjected to periodic boundary conditions. 
Potential energy $V_\text{pot}$ (top) and lateral force $F_\text{\text{tot}}$ (center)
as a function of the moved distance $x$
at a small sliding velocity of $v = 10^{-5}$.
The bottom panel shows the displacements of atoms on the abscissa and the atom number on the ordinate
at three different displacements. 
The dashed lines mark the mean lateral force (center panel) and 
the center-of-mass displacement with the same color as the displacements. 
}
\end{figure}

The motion of neighboring beads gets increasingly correlated with increasing elastic coupling.
While the chain behaves like a stiff rod for very large $k$, substantial multi-stability prevails at medium $k$, which means that there can be more than one mechanically stable microscopic configuration for a given center-of-mass position. 
Such multistability is at the root for plasticity and Coulomb friction, as discussed in an early work by Prandtl~\cite{Prandtl1928ZAMM,Popov2012ZAMM}.
Here, we refer to medium coupling when the coupling is still small enough for substantial multistability to exist but large enough for the width of a kink or discommensuration to be of order unity. 
For medium coupling, we primarily consider a value of $k = 10 V_0/a^2$, which leads to a kink (half) width of 
\begin{equation}
\zeta = a \sqrt{k^*}
\end{equation}
in the continuum approximation of the FK model, which is also known as the sine Gordon model. 
This expression evaluates numerically to $\zeta = 0.53~a$, which is reasonably close to the value of $\zeta = 0.63~a$ that was deduced through fits to data like that shown in the bottom panel of Fig.~\ref{fig:barr10pbc}.

Fig.~\ref{fig:barr10pbc} shows the dynamics of an athermal chain subjected to periodic boundary condition that occurs when it is moved at a small, constant center-of-mass velocity relative to the corrugated potential. 
The bottom panel shows the actual phase shifts on the absicca and the bead or atom numbers on the ordinate. 
The kink, which is enforced via periodic boundary conditions, moves through the chain. 
Each time, an atom in the chain passes through an energy maximum, it quickly advances to the next available energy minimum, once its old position turns from marginally stable to unstable. 
The total potential energy at depinning is roughly $V_\text{pot}^\text{dep} \approx 4.2003~V_0$ from where it falls into the next available minimum at $V_\text{pot}^\text{nxt} \approx 3.4394~V_0$.
This makes the energy drop be $\Delta E_\text{d} = V_\text{pot}^\text{dep} - V_\text{pot}^\text{nxt} \approx 0.7608\,V_0$.
This quantum equals the energy that is dissipated each time the kink advances by one substrate period.
The absolute energy minimum is only marginally lower for the given chain, namely
$V_\text{pot}^\text{min} = 3.4223~V_0$, which leads to an upper bound for the Peierls-Nabarro barrier of 
$\Delta E_\text{PN}^\text{max} = V_\text{pot}^\text{dep} - V_\text{pot}^\text{min} \approx 0.7786~V_0$. 
%
%
This latter value is in good agreement with the analytical expression for the barrier height in the strong-coupling approximation
\begin{equation}
\label{eq:barrier_analytical}
\Delta E = 16 k b^2 e^{-\pi^2\sqrt{k^*}},
\end{equation}
which evaluates numerically to $\Delta E = 0.9101~V_0$. 
Our value is an upper bound for the barrier since the center of mass position is not necessarily a good reaction coordinate.
Paths with energy barriers can (and do) exist that are lower than those encountered when stepping  the center-of-mass forward adiabatically without thermal fluctuations. 
Thus, the current coupling lowers the energy barrier for kink motion to less than about 40\% of what it would be without coupling, which is $2V_0$.
%

At low temperatures, the true Peierls-Nabarro barrier $\Delta E_\text{PN}$ will determine the kink mobility, which in turn determines that of the chain and thereby the equilibrium viscosity.
As a result, $\eta_0$ is expected, to leading order, to obey an Arrhenius dependence with $\eta_0 \propto \exp(-\beta \Delta E_\text{PN})$.
%
%
In contrast, the friction at medium to high velocity is dictated by $\Delta E_\text{d}$.
This is because chains undergo enforced basin hopping at high velocities.
Their dynamics then resemble the athermal dynamics at small velocities, unless the temperature is high.  
Since $P$ instabilities occur when the chain's center of mass is advanced by the substrate's lattice constant an athermal kinetic friction force of $f_\text{k} = \Delta E_\text{d}/b$ per atom ensues, which evaluates numerically to $f_k \approx 0.7133~V_0/a$, which is essentially the same result, as when dividing the  total kinetic friction force of $F_k \approx 11.414$ by 16, the number of atoms in the chain.
Thus, when the chain moves too quickly for the instabilities to be triggered prematurely through thermal fluctuations, an effective viscosity of $\eta_\text{eff} \approx \Delta E_\text{d}/(vb)$ is expected.

\begin{figure}[htbp]
\includegraphics[width=0.9\columnwidth]{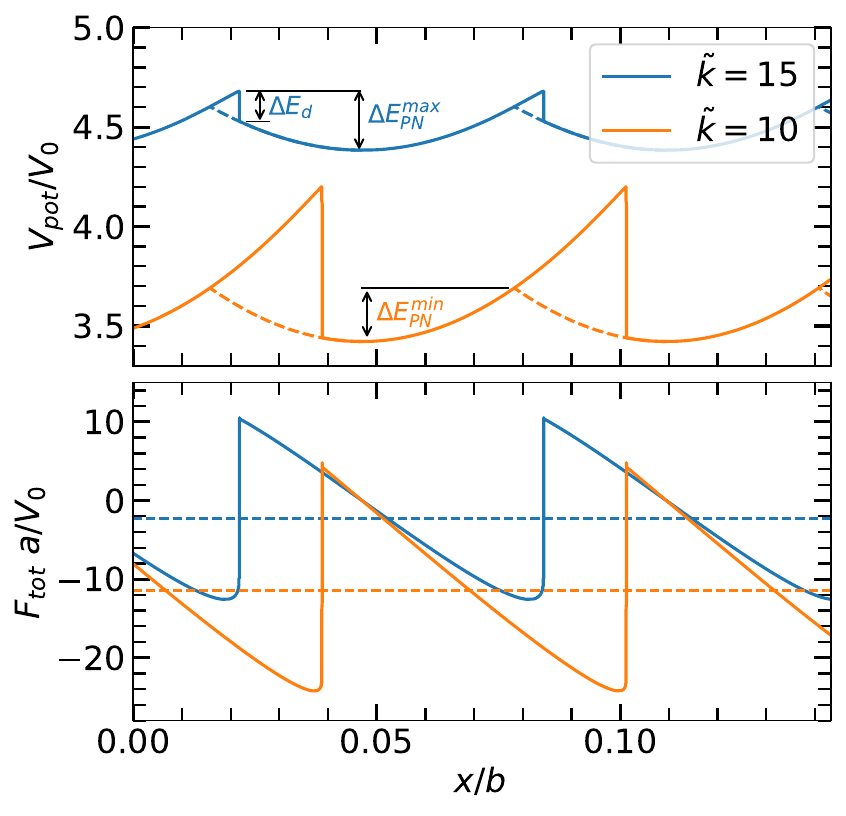}
\caption{\label{fig:barr_stiff_k10_15_pbc}
Evolution of the total potential energy $V_\text{pot}$ (top) and total lateral force $F_\textrm{lat}$ (bottom) as a function of a moving center-of-mass position $x$ for a $k = 15~V_0/a^2$ and a $k = 10~V_0/a^2$ chain at $k_\text{\text{B}}T = 0$. 
For the $\tilde{k} = 15$ chain, the energy drop $\Delta E_\text{d}$ and bounds for the Peierls-Nabarro barrier $\Delta E_\text{PN}$ are shown. The upper and lower bounds of $\Delta E_\text{PN}$ for $\tilde{k}=10$ are $0.778~V_0$ and $0.267~V_0$ respectively; for $\tilde{k}=15$ they are $0.297~V_0$ and $0.218~V_0$.
}
\end{figure}

To corroborate our claims, the dynamics of a $k = 15~V_0/a^2$ chain is considered in addition to that of the former model, see Fig.~\ref{fig:barr_stiff_k10_15_pbc}, where the basin hopping process of that chain type is depicted in detail and contrasted to that of the default model.
This time, $\Delta E_\text{d} = 0.153~V_0$ is merely half of $\Delta E_\text{PN} = 0.297~V_0$.
Thus, the lateral force per atom deduced from $\Delta E_\text{d}$ is $f_\text{k} = 0.1432~V_0/a$.
This ratio is again very close to that deduced from the direct measurement shown in Fig.~\ref{fig:barr_stiff_k10_15_pbc}, i.e., $2.2941 / 16 = 0.1434$.
To achieve this level of agreement, sliding velocities of $10^{-5}$ were employed. 
%

%
%
%


Figure~\ref{fig:barr_stiff_k10_15_pbc} also shows a lower-bound estimate for the Peierls--Nabarro barrier. This estimate is based on the system’s energy when the center of mass reaches the transition point — a position where multiple microscopic configurations are possible. One configuration is assumed before an instability occurs, while the other would be realized if the motion were reversed after the instability has taken place.
These two configurations share the same center-of-mass position but cannot be transformed into one another without an energy cost. That energy difference constitutes the missing contribution to the full Peierls--Nabarro barrier.

\subsubsection{Thermal dynamics of the default model}
\begin{figure}[htbp]
\includegraphics[width=0.9\columnwidth]{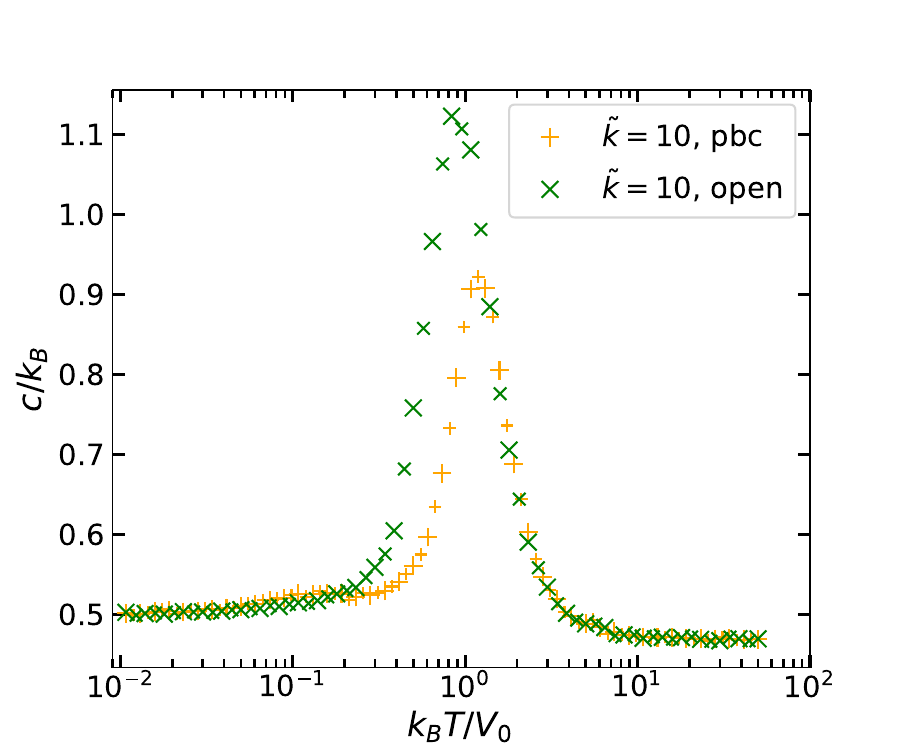}
\caption{\label{fig:cp_10k}
Configurational
specific heat $c_\text{c}$ per degree of freedom as a function of reduced temperature $k_BT/V_0$ for free atoms Frenkel Kontorova chains with medium strong springs, $\tilde{k} = 10$, which can be open (green crosses) or subjected to periodic boundary conditions (pbc, orange plus symbols). 
Larger symbols are deduced from finite difference of $U(T)$, smaller symbols from energy fluctuations.
}
\end{figure}

\begin{figure}[htbp]
    \centering
    \includegraphics[width=0.9\columnwidth]{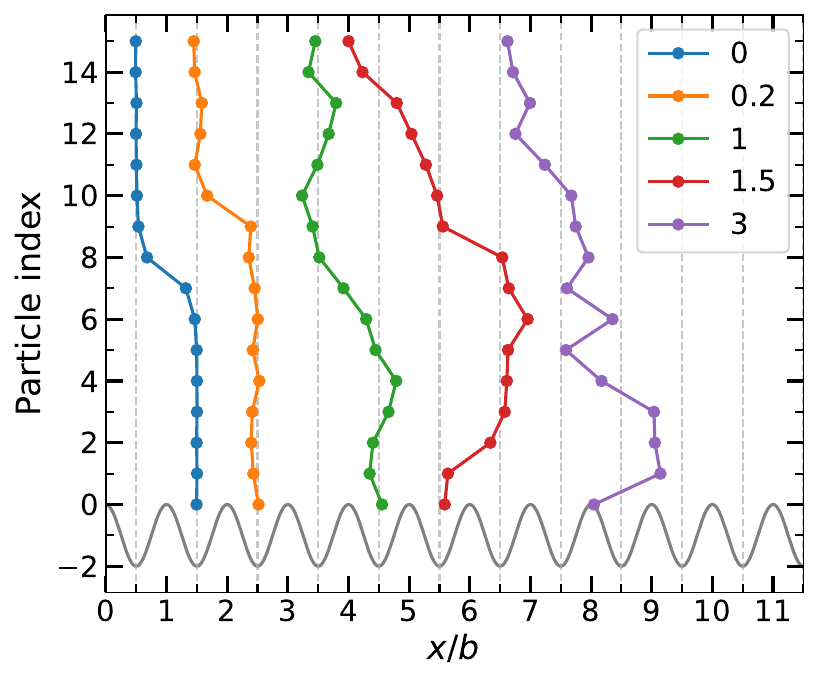}
    \caption{Representative configurations of the $\tilde{k}=10$ chain with periodic boundary conditions at different reduced temperatures $\tilde{T}$. The substrate potential is shown in gray for reference only. Vertical grid-lines pass through its minima.
    }
    \label{fig:configs_temp_k10}
\end{figure}

We note that the finite elastic coupling between the beads does not alter the dependence of specific heat on temperature qualitatively compared to the uncoupled case.
However, the location of the peak maximum, $T^*$, moved to a temperature about 2.5 times higher than for very weak coupling (Fig.~\ref{fig:cp_10k}), although the energy barrier to be passed was reduced by a factor of two.
In the medium-coupling limit, the peak in specific heat arises from the thermal activation of structural excitations, i.e., near and above $T^*$ more than one discommensuration is likely to occur.
Thus, there is a non-negligible occurrence of one or several additional pairs of kinks and anti-kinks in addition to the kink enforced by the periodic boundaries.
Representative chain configurations at different temperatures are depicted in Fig.~\ref{fig:configs_temp_k10}.
A super-Arrhenius mobility of the chain due to an increased number of defects is the natural consequence of these results. 

The number of kinks at a safe distance below $T^*$ is usually unity in our periodically repeated chain.  
At this temperature, the Peierls-Nabarro barrier is also greater than the thermal energy, which turns the undriven dynamics into a process where the diffusion of a kink is thermally activated.  
This implies that kink motion occurs via rare thermally activated hops over energy barriers rather than continuous sliding.
As a consequence, the transition from ballistic motion to diffusion must pass through an intermittent regime.  
In this regime, the kink motion alternates between being trapped in local energy minima and sudden jumps, resulting in nontrivial temporal correlations and anomalous transport properties.
For the periodically repeated $\tilde{k}=10$ chain, this happens via sub-diffusive motion, where the diffused squared distance $\Delta x^2$ satisfies  
\begin{equation}  
\left\langle \left\{ x(t) - x(0) \right\}^2 \right\rangle \propto t^\alpha  
\end{equation}  
with $0 < \alpha < 1$.  
Such sub-diffusive behavior is characteristic of complex liquids~\cite{Metzler2000PR}.  
The simplest \emph{atomistic} model system having revealed sub-diffusive dynamics so far is arguably a (supercooled) binary Lennard-Jones mixture~\cite{Kob1995PRE}.

\begin{figure}[htbp]
\includegraphics[width=0.9\columnwidth]{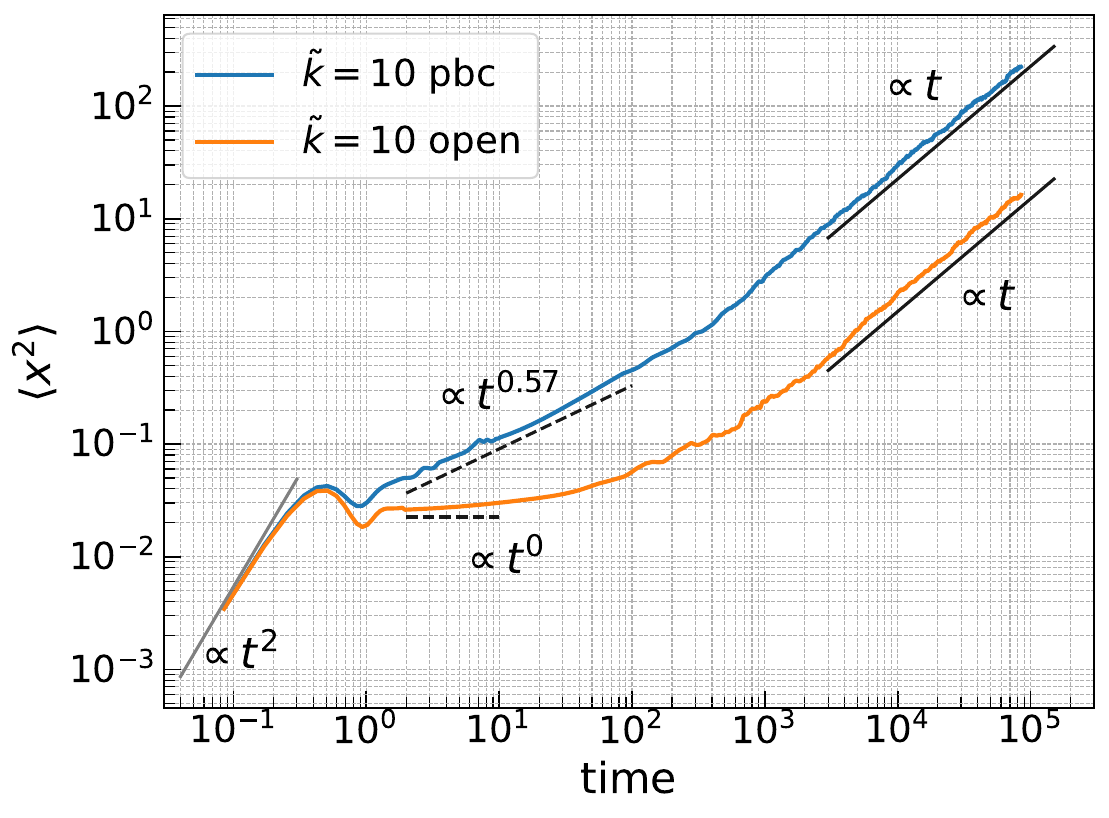}
\caption{\label{fig:diff_bc}
RMS displacement dynamics for a $\tilde{k}=10$ chain at $\tilde{T} = 0.5$ under periodic and open boundary conditions. The gray line represents the initial super-diffusive regime with power-law of 2, the black dashed line indicates sub-diffusive behavior, and the solid black line corresponds to the diffusive regime with power-law  of 1.
}
\end{figure}

While the driven dynamics of the \emph{open} $\tilde{k}=10$ chain is discussed in a separate section, it is appropriate to discuss the thermal dynamics here.
Its sub-diffusive exponent has dropped to less than 0.1 compared to $\alpha \approx 0.57$ for the periodic chain, and its ultimate diffusion is substantially reduced, as shown in Fig~\ref{fig:diff_bc}.
This can be attributed to edge effects in the open chain, which allow it to sink more deeply into energy minima, resulting in more pronounced sub-diffusion than in the periodic chain.
The absence of edges in the periodic chain enables collective center-of-mass motion, which helps overcome barriers more effectively, whereas the open chain’s broken translational symmetry leads to stronger local pinning and suppressed coordinated motion.
\subsubsection{Driven dynamics of the default model}

\begin{figure}[h]
    \centering
    \includegraphics[width=0.9\linewidth]{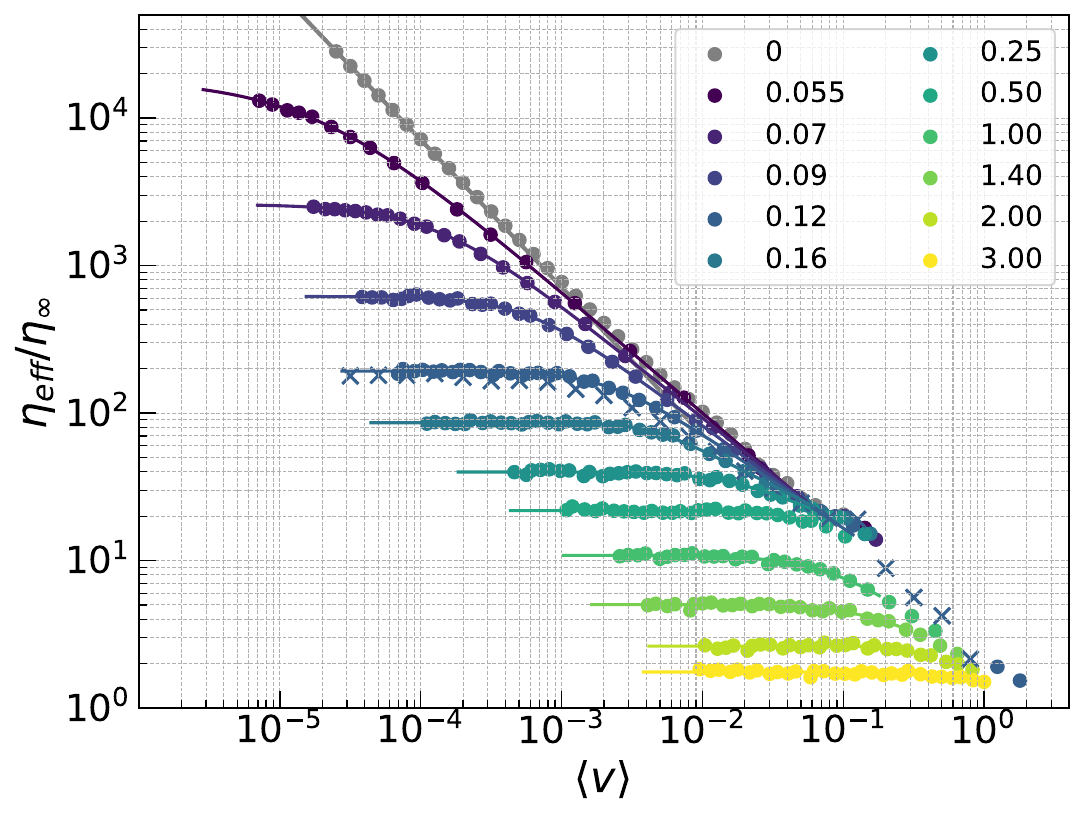}
    \caption{Effective viscosity of a $\tilde{k}=10$ chain with PBC in a corrugated potential. 
     Circles represent data obtained from simulations at constant force and reduced temperature $\tilde{T}$.
     For $\tilde{T} = 0.12$, chains were driven at a constant center-of-mass velocity.
     Lines reflect fits to the Carreau-Yasuda equation, except for the grey line, showing $\Delta E_d/bv$ per atom.
     %
     }
    \label{fig:eff_visc_k10_CarrYas}
\end{figure}

Fig.~\ref{fig:eff_visc_k10_CarrYas} shows results for the effective viscosity of the $k = 10~V_0/a^2$, $N = 16$, and $a/b = 15/16$ chain.
The rheological response resembles, as is the case for uncoupled atoms, that of many non-Newtonian liquids: the equilibrium viscosity $\eta_0$ increases with decreasing temperature and the effective viscosity can be described well with the Carreau-Yasuda (CY) equation,
\begin{equation}
\eta = \frac{\eta_0}{\left( 1 + (\dot{\gamma}/\dot{\gamma}_0)^a\right)^{(1-n)/a}},
\end{equation}
which in addition to the equilibrium viscosity $\eta_0$ and a characteristic shear rate $\dot{\gamma}_0$ depends on the shear-thinning exponent $n$ and the Yasuda exponent $a$.
Not only CY but also the Eyring equation fits the $\eta_\text{eff}(\dot{\nu})$ dependences shwon in Fig.~\ref{fig:eff_visc_k10_CarrYas} very well again, in particular for thermal energies less than $V_0$.
This is in contrast to the Prandtl model, the mean-field variant of the Frenkel-Kontorova model, where CY clearly outperforms Eyring~\cite{Gao2023TL}.

A few observations in Fig.~\ref{fig:eff_visc_k10_CarrYas} are worth discussing.
First, the ratio $\Delta E_\text{d}/(bv)$ forms indeed a good upper bound or even estimate for the true effective viscosity, the more so, the deeper the temperature.
Second, at high velocity and low temperature, there is a gap in the values of possible velocities, when the chains are driven under a constant force.
This gap can be attributed to a shear-thinning exponent that would hypothetically fall below the value of $n = 0$ in the range of forbidden velocities, which implies unstable motion.
In fact, when forcing the center-of-motion to lie in this unstable range, friction decreases with increasing velocity in that range. 
This is the sign for the coexistance of two running solutions in the Frenkel-Kontorova model~\cite{Braun1997PRE}, which are well known.
They take a peculiar form in the quantum Frenkel-Kontorova model, where a tunneling solution can coexisting with a running solution~\cite{Krajewski2004PRL}.  
Third, outside the regime of dynamical bistability, the effective viscosity does not depend much on whether a constant velocity or a constant force is imposed.
%
%
%

\begin{figure}[htbp]
\includegraphics[width=0.9\columnwidth]{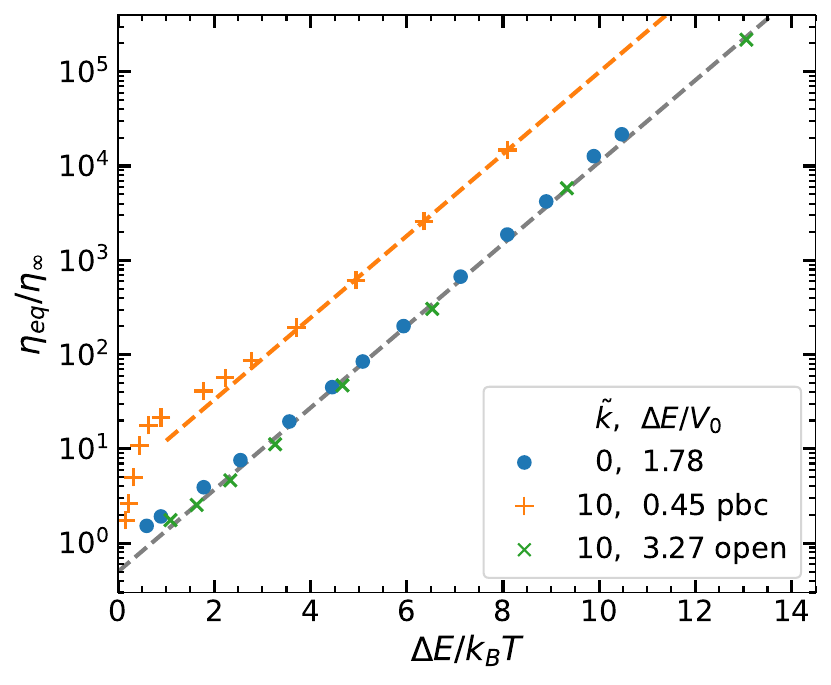}
\caption{\label{fig:eq_visc_k10}
Equilibrium viscosity as a function of temperature for free atoms (blue circles).
Data for medium coupled ${\tilde{k}}= 10$, is included, 
one time for periodic (orange plus symbols) and one time with open (green crosses) boundary conditions is included. 
Lines with a slope of one are drawn to guide the eye. 
}
\end{figure}

At medium coupling, the temperature dependence of $\eta_0$ is strongly altered compared to the limit of uncoupled atoms, as can be seen in Fig.~\ref{fig:eq_visc_k10}.
The apparent activation energy $\Delta E_\text{app} = - \partial \ln \eta / \partial \beta$ with $\beta = 1/k_B T$ first increases with decreasing temperature at very high temperatures, in a similar way as the free atoms.
However, $\Delta E_\text{app}$ drops to a substantially smaller value in the vicinity of $T^*$ and becomes quite constant at even lower temperatures. 
At the lowest investigated thermal energy, $k_BT \approx V_0/20$, it assumes a value of $\Delta E_\text{app} \approx 0.5\,V_0$, which is in between the lower and upper bound given by the potential of mean force and the barrier for athermal sliding, respectively. 
%
The behavior is somewhat reminiscent of some glass forming melts, which can also undergo a transition from non-Arrhenius to Arrhenius behavior near a temperature, at which the specific heat assumes a local maximum~\cite{Angell1995S}.

\begin{figure}[htbp]
\includegraphics[width=0.9\columnwidth]{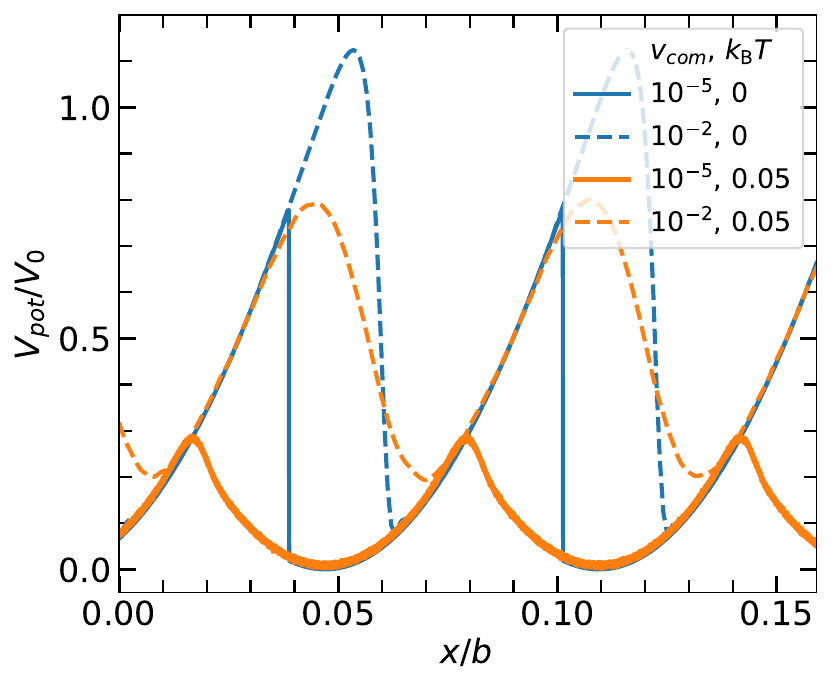}
\caption{\label{fig:barr_temp_k10_pbc}
Athermal and thermal energy barrier for medium coupling ($\tilde{k} = 10$), for two center-of-mass sliding velocities using periodic boundary conditions. For clear visualization, the curves for $\tilde{T}=0$ are shifted in y-direction by 3.42, while that for $0.05$ by 3.79.
%
%
}
\end{figure}

It remains to be understood why the effective viscosity at high sliding velocities and high temperatures can be deduced so accurately from the slow-velocity dynamics at low temperature.
To this end, we contrast the mean instantaneous energy as a function of slid distance for both cases in Fig.~\ref{fig:barr_temp_k10_pbc}.
One can see that the process in the fast thermal chain {(shown in dashed orange)} is merely a delayed, smeared-out variant of the slow, athermal chain  dynamics {(shown in solid blue)}.
The instabilities occur later and at a higher energy, as they have less time—or rather, less distance—to develop than in the slow, athermal case.
However, the energy drop also occurs to a higher level than before so that the difference, which is the dissipated energy, remains roughly identical.
Fig.~\ref{fig:barr_temp_k10_pbc} also reveals that the barrier deduced from the (internal) energy for the slowly moving system is similar to that deduced from Fig.~\ref{fig:barr_stiff_k10_15_pbc}.
Both barriers are only lower bounds for $\Delta E_\text{PN}$, since a typical structure left of the barrier cannot be continuously deformed into one to the right of the barrier.

\subsubsection{Non-uniform force distribution and constraints}

\begin{figure}
    \centering
    \includegraphics[width=0.9\linewidth]{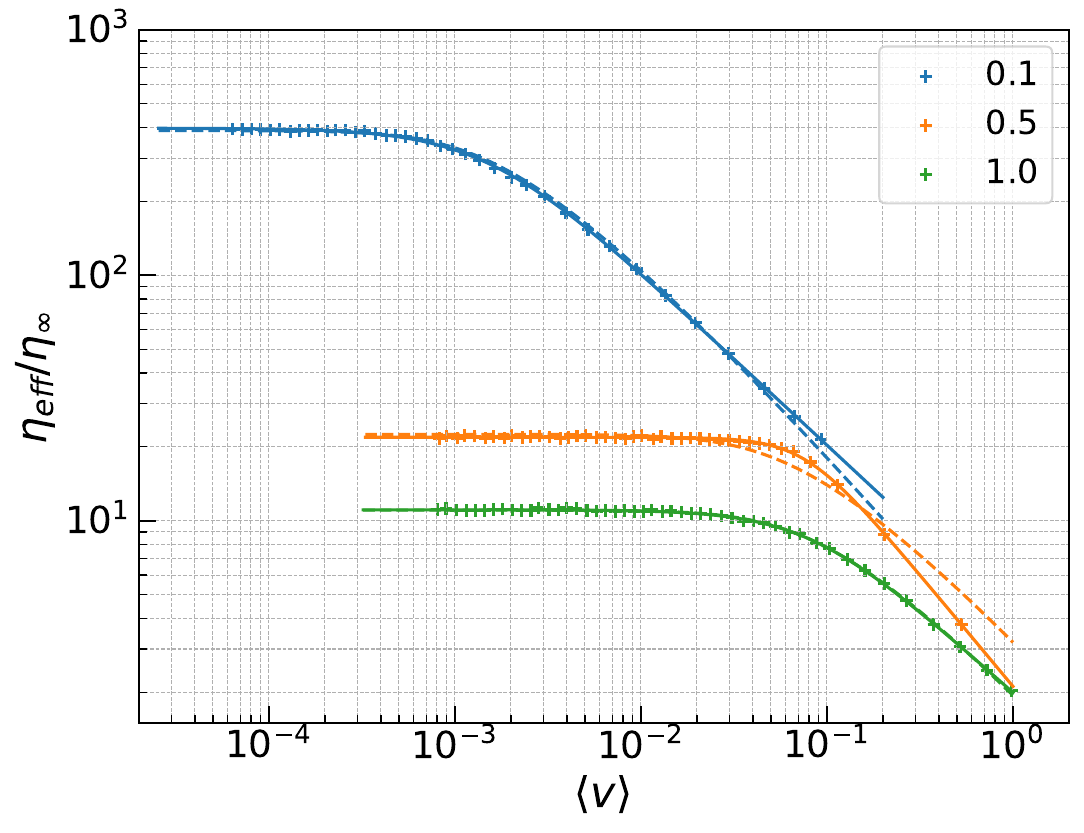}
    \caption{Effective viscosity of a $\tilde{k}=10$ chain with  periodic boundary condition and force applied to alternate atoms for three temperatures. Eyring fit is shown in dashed while Carreau-Yasuda fit is shown by solid lines.}
    \label{fig:fk_as_prandtl}
\end{figure}

Throughout most of this work, the external driving force is distributed uniformly among the beads in the chain.
This can be said to correspond to a body force. 
In reality, external forces do not directly apply to atoms or molecules in a sheared liquid, but indirectly via intermittent layers.
Thus, the external force applied to an atom or molecule wanting to move past a barrier provided by an adjacent lamella is not a constant of time, but gets continuously ramped up to a maximum such that the effect of the external force is non-uniform in space and time. 
To explore such an effect, we also studied a chain, in which the external force was applied only to odd-numbered beads.
This, however, does not break the validity of the Eyring model for the FK chain, at least not at lower temperature, as can be seen in Fig.~\ref{fig:fk_as_prandtl}.
It also leaves the equilibrium values unchanged: the (reduced) equilibrium viscosity of $\tilde{\eta}_\text{eq}(\tilde{T}=0.1) \approx 400$ found in Fig.~\ref{fig:fk_as_prandtl} corresponds to that shown in Fig.~\ref{fig:eq_visc_k10}, where $\Delta E/k_BT = 0.45 / 0.1 \to 4.5$. 

\begin{figure}
    \centering
    \includegraphics[width=0.9\linewidth]{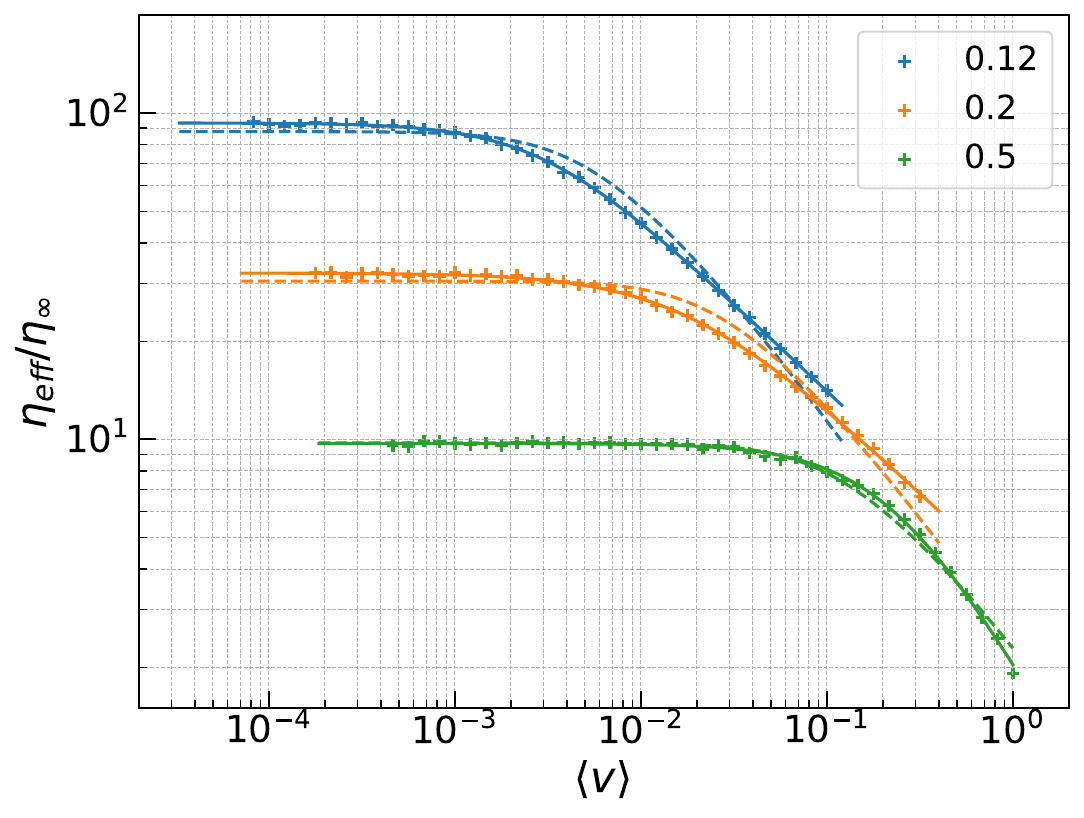}
    \caption{Effective viscosity of a $\tilde{k}=10$ chain with  periodic boundary condition and constant velocity applied to alternate atoms for three temperatures. Eyring fit is shown in dashed while Carreau-Yasuda fit is shown by solid lines.}
    \label{fig:fk_as_real_prandtl}
\end{figure}

{However, the equilibrium viscosity as well as the functional form of the rate dependence clearly change when the velocity of every other bead is constrained to $v_0$, as shown in Fig.~\ref{fig:fk_as_real_prandtl}.
This constraint, which might pertain to selectively actuated materials \cite{Sim2025AFM,De2005BPh,Vos2024COCB}, creates a microscopically sheared system akin to a multi-particle Prandtl model, for which Carreau-Yasuda outperforms Eyring equation \cite{Gao2023TL}.  } 

\subsubsection{Alternative damping}

\begin{figure}
    \centering
    \includegraphics[width=0.9\columnwidth]{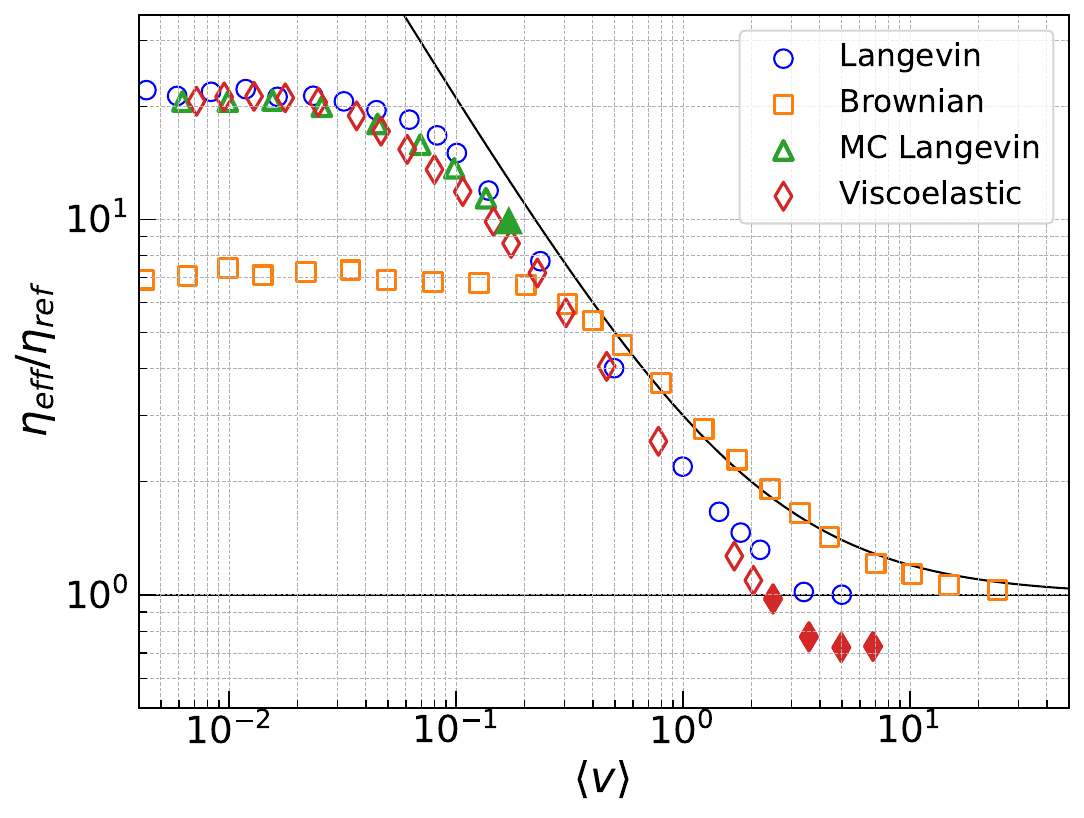}
    \caption{Effect of dynamical properties and thermostats on rheological behavior for $\tilde{k}=10$ chain at $\tilde{T}=0.5$. The (local) time constant for both underdamped Langevin (solid circle) and momentum-conserving Langevin (solid triangle) is unity. The latter becomes unstable for higher velocities and viscosity drops at much higher rate shown by dashed green line.  Damping coefficient for overdamped Langevin (Brownian) is 1.0. Viscoelastic Langevin has a Maxwell element of stiffness $0.5k$ and damping coefficient $0.2m/\tau$ in parallel to the dashpot of damping coefficient $0.8m/\tau$. The slanted black line has a slope of $-1$, the lower limit for stable solution.}
    \label{fig:fk_therm_effect}
\end{figure}

From our analysis conducted so far, shear thinning has been rationalized from the perspective of basin hopping.
In this view, the rate at which excess kinetic energy—produced by instabilities—is dissipated via the damping term is assumed to play a subordinate role.
This assumption may be justified under certain conditions, particularly at intermediate sliding velocities, which are large enough so that thermal activation takes too long to matter but small enough that the degrees of freedom have enough time to dissipate the kinetic energy, which was released during the last instability, before the next one occurs. 
Fig.~\ref{fig:fk_therm_effect}, which shows the effective viscosity of the $\tilde{k}=10$ chain under different dynamical schemes, confirms this expectation.
In the following, we attempt to rationalize how various damping schemes and thus thermostats influence the shear thinning behavior.

For Brownian dynamics, the same damping $\gamma = m/\tau$ was chosen as for Langevin dynamics.
However, the Brownian particle does not have a mass.
A smaller mass implies a larger attempt frequency for barrier crossing, so that the Brownian chain exhibits faster diffusion, which makes it have a smaller equilibrium viscosity than the Langevin chain.
At large, but not extremely large, sliding velocities, the massless Brownian atoms more easily sink into energy minima, as their (nonexistent) kinetic energy cannot assist in overcoming the next barrier.
As a result, they become more deeply trapped in energy minima than their Langevin counterparts.
Consequently, the Brownian chain exhibits higher friction than the Langevin chain at high velocity and follows the theoretical estimate $\eta \approx \Delta E / v + \eta_\infty$ more closely over a wider range of velocities.
At extremely large velocities, the Brownian particle can no longer relax into any energy minimum so that Brownian and Lagrangian FK chains are damped with $\eta_\infty$, which for the Brownian and regular Langevin FK chains defines $\eta_\textrm{ref}$ in Fig.~\ref{fig:fk_therm_effect}.

When atoms in the FK chain only experience relative damping but none with respect to the substrate, $\eta_\infty$ vanishes.
This is because the relative motion between adjacent atoms in the chain is rather minor at very large sliding velocity, which leads to small and ultimately vanishing dissipation.
In fact, the shear-thinning exponent can drop below zero so that stable motion under a constant force above a critical value is no longer possible.
This is why the pertinent $\eta(\dot{\gamma})$ data, reflected by the full green triangles in Fig.~\ref{fig:fk_therm_effect}, terminate at a critical velocity $v_\textrm{c}$.
Data above $v_\textrm{c}$, which is shown using open symbols, was collected using a constant center-of-mass velocity constraint rather than constant force.
For completeness, we note that the reference $\eta_\infty$ was chosen such that $\eta(\dot{\gamma})/\eta_\infty$ overlapped with the default Langevin chain, even though the measured damping in absolute terms was smaller than that of the Brownian chain.

The final dynamical model explored for the periodic $\tilde{k} = 10$ chain is the viscoelastic model, where damping is relative to the substrate but occurs, in part, with a delay.
While the overall behavior is not substantially affected by the specific choices made,
the range where $\eta_\textrm{eff}$ scales roughly inversely with $\langle v \rangle$ is extended compared to the case with instantaneous damping.

\subsection{Open FK chain at medium elastic coupling}

\begin{figure}[thbp]
\includegraphics[width=0.9\columnwidth]{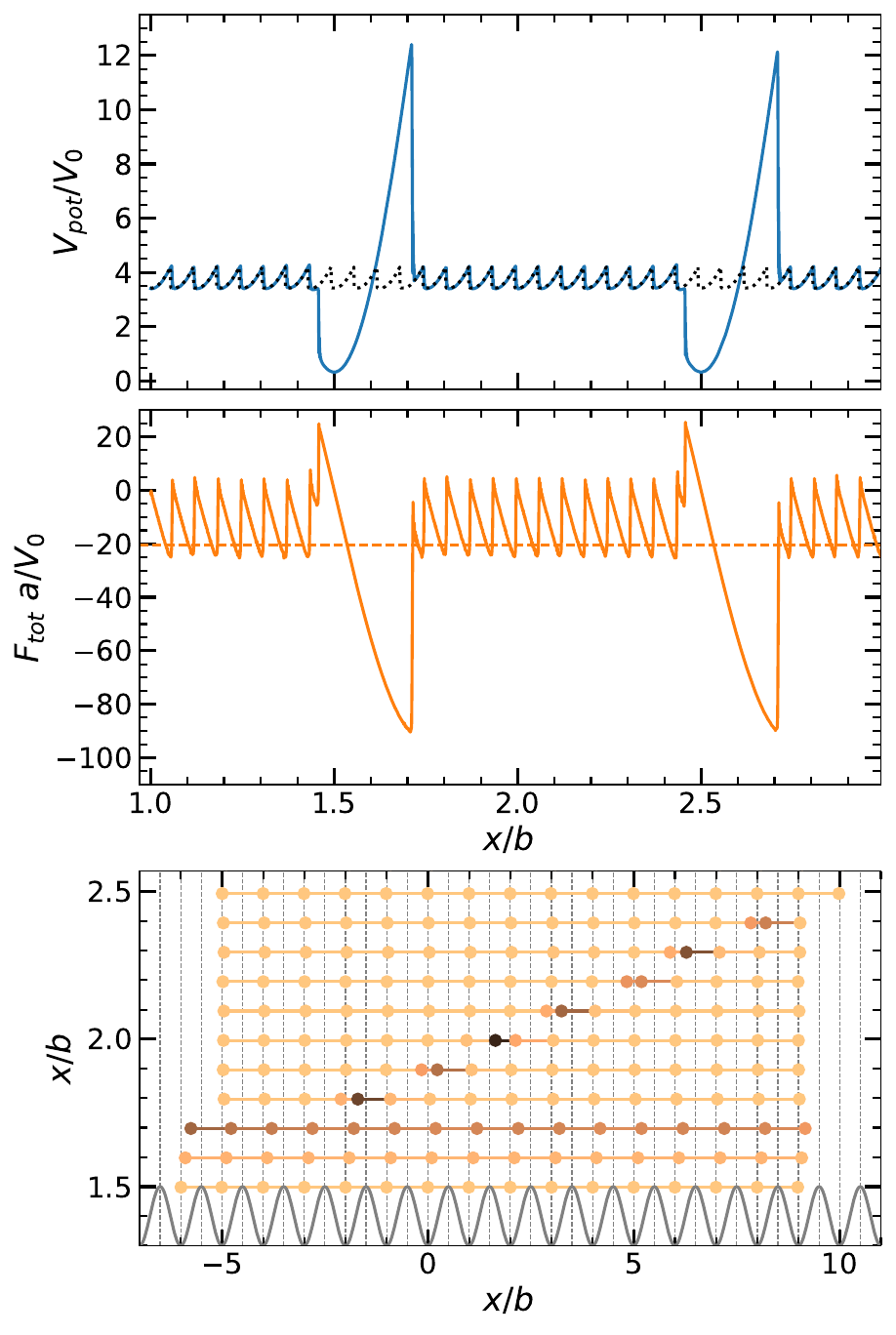}
\caption{\label{fig:barr10open}
Similar to Fig.~\ref{fig:barr10pbc} but this time for a $\tilde{k} = 10$ chain with open ends. 
The full lines are fits to Carreau-Yasuda in the top panel and to Eyring in the bottom panel.
The thin dotted in the top panel shows the dynamics of a periodically repeated chain. The bottom panel illustrates the dynamics of discommensuration with the chain movement. The positions of individual atoms is shown on the abscissa while their mean is shown on the ordinate.
}
\end{figure}

So far, we have focused on periodically repeated chains, which enforces the existence of at least one discommensuration.
Releasing that constraint allows the open chain to adopt a more favorable ground state, whereby it can sink into a substantially deeper energy minimum than the periodic chain. 
This in turn increases the barrier to advance the chain by a substrate period. 
When translating the open chain, inequivalent instabilities occur.
Starting from a kink-free ground state, where all atoms line up in the same  minimum of the substrate potential minima, e.g., $x/b \approx 1.5$ in Fig.~\ref{fig:barr10open}, the chain remains kink-free up to the instability point at $x/b \approx 1.7$.
At that point, a discommensuration pops in, which constitutes the most dramatic energy drop $\Delta E_d \approx 8.4~V_0$. 
After pop-in, the discommensuration is located between the chain's trailing end and its center of masses.
Now, the motion of the open chain resembles that of a chain with periodic boundary condition, i.e., each time the chain advances by $b/P$, the discommensuration advances by one lattice constant.
These instabilities are rather minor individually but sum up to an accumulated energy drop of $\Delta E_\textrm{d} \approx 10.44~V_0$ over 12 events.
Finally, there is a pop-out of the discommensuration with an associated $\Delta E_\textrm{d} = 2.4~V_0$, after which the sequence repeats.
These dynamics, which the open chain undergoes under a small, constrained center-of-mass motion reveals once more that the athermal Peierls-Nabarro barrier can be a poor indicator of the true barrier, since, the discommensuration can thermally nucleate at the end of the chain given enough time.

Assuming that, at intermediate velocities and low temperatures, dissipation is governed more by the energy landscape than by the system's dynamical properties sets the expectation that the friction in this regime is approximately given by the total sum over elementary energy drops $\Delta E_\text{d}^{(n)}$ per period divided by the period $b$.
In fact, the effective viscosity computed using this hypothesis gives a reasonably accurate hull function, as demonstrated in Fig.~\ref{fig:eff_visc_k10_obc}, where the hull covers an increasingly large domain with decreasing temperature.
At those points where the hull is a good approximation to the true data, one may thus argue that the dynamical properties, such as mass or damping, are of relatively minor importance—at least within a certain parameter window.
In other words, a good fit of some rheological model, in which prefactors are defined by damping or the square root of the mass, would not indicate the validity of the underlying theory, but rather result from more fortuitous reasons.

\begin{figure}[h]
\includegraphics[width=0.9\columnwidth]{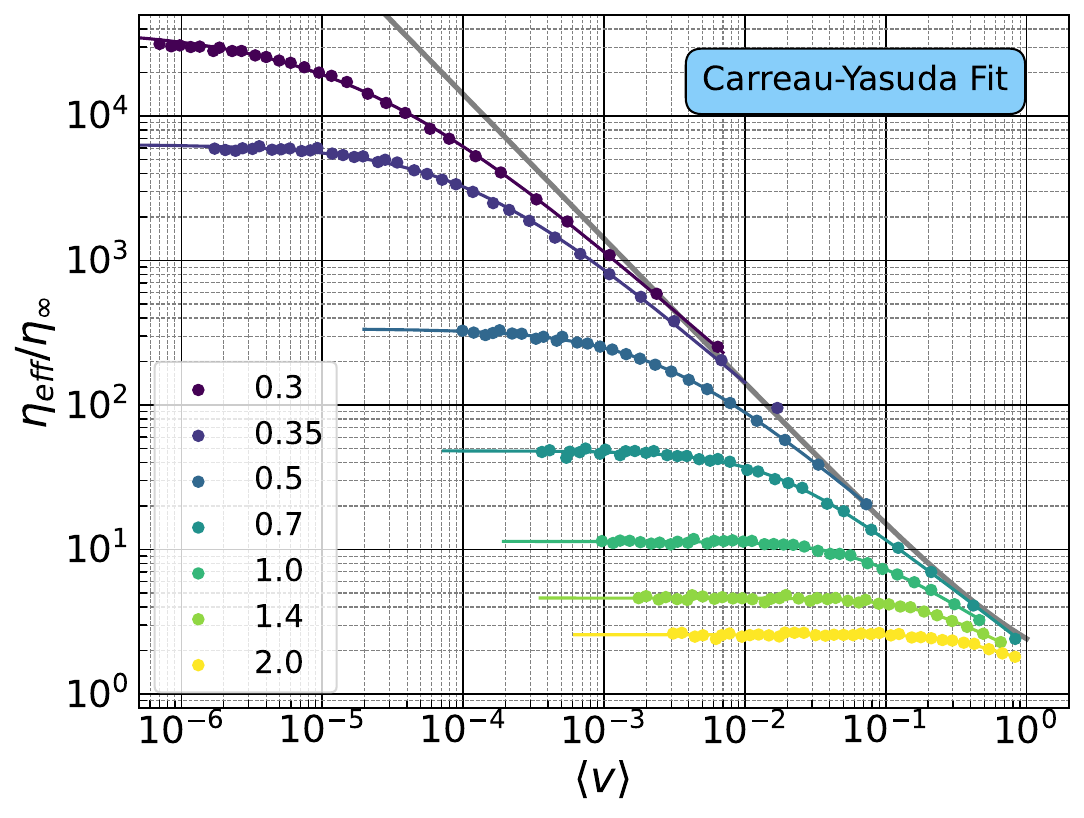}
\includegraphics[width=0.9\columnwidth]{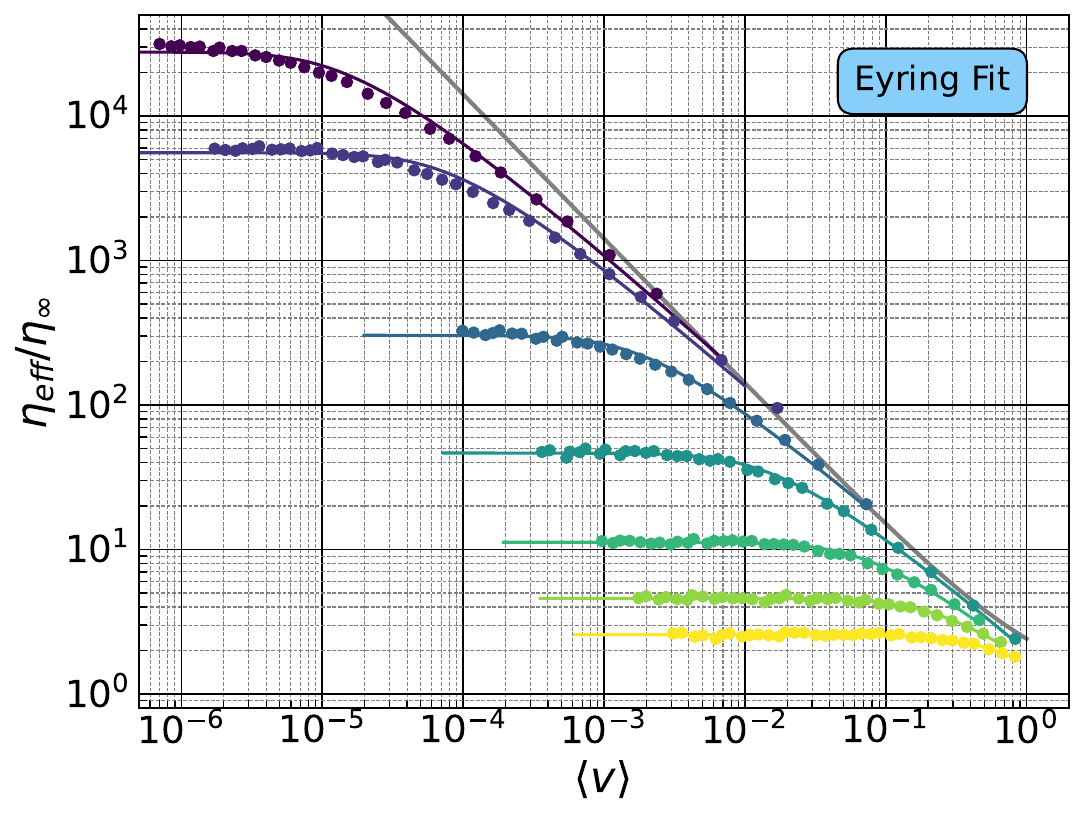}
\caption{\label{fig:eff_visc_k10_obc}
Effective viscosity of a $\tilde{k}=10$ open chain in a corrugated potential. 
Circles represent data obtained from simulations at constant force and reduced temperature $\tilde{T}$.
Lines reflect fits to the Carreau-Yasuda (top) and Eyring (bottom) equations, except for the gray line, showing $\sum_n E_d^{(n)}/bv$  per atom. 
}
\end{figure}

Although Eyring theory provides a seemingly accurate fit to the data, it slightly but noticeably underestimates the viscosity at velocities where shear thinning becomes apparent when the temperature is low.
This appears to be a general phenomenon, whose observation, however, requires high data precision and proper fitting—i.e., fitting the logarithm of viscosity rather than viscosity itself.



%
     %
     %

\subsection{FK chains at strong elastic coupling}

\begin{figure}[htbp]
\includegraphics[width=0.9\columnwidth]{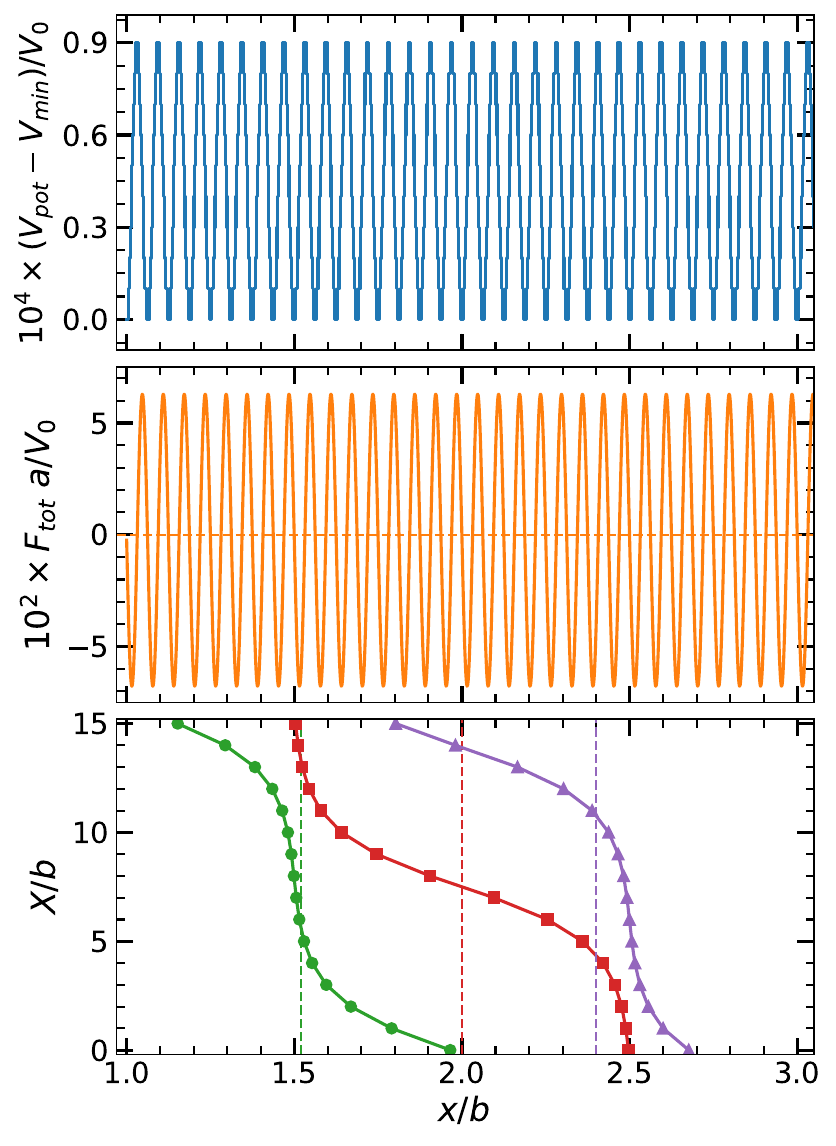}
\caption{\label{fig:barr100_pbc}
Athermal dynamics of a $\tilde{k} = 100$ chain subjected to periodic boundary conditions. 
Potential energy $V_\text{pot}$ (top) and lateral force $F_\text{lat}$ as a function of the moved distance $x$
at a small sliding velocity of $v = 1e-5$.
The top panel shows the displacements of atoms on the abscissa and the atom number on the ordinate
at three different displacements. 
The center-of-mass displacement is marked by dotted lines having the same color as the displacements. 
}
\end{figure}

When the elastic coupling increases, discommensurations extend over larger domains, as is depicted in Figs.~\ref{fig:barr100_pbc} for $\tilde{k} = 100$, which is the value used throughout this section. 
The advancement of a discommensuration now entails a rather minor transition barrier of $\Delta E = 9.000 \times 10^{-5}~V_0$, which is in very good agreement with the continuum theory, Eq.~\eqref{eq:barrier_analytical}, yielding $9.627 \times 10^{-5}~V_0$.
The enhanced rigidity counteracts the sudden release of energy when a discommensuration advances by one lattice constant, which suggests that the chain can glide past the substrate without instabilities and thereby allow the chain to be driven (quasi-) adiabatically at very low temperature. 
%

The stiff, periodic chain turns out to be the only model with a marginal anomaly in the specific heat at a thermal energy of order $V_0$.
Its peak even barely exceeds $k_B/2$, which is the (classical) specific heat just above $T = 0$, when all modes are harmonic and not only the internal modes. 
At small (but not extremely small) temperature, the specific heat assumes a value in the immediate vicinity of $c_\textrm{c}/k_B = (1-1/P)/2$, which implies that the chain cannot be trapped for a long time near a harmonic minimum.
Thus, the center of mass can be expected to be diffusive. 

\begin{figure}[htbp]
\includegraphics[width=0.9\columnwidth]{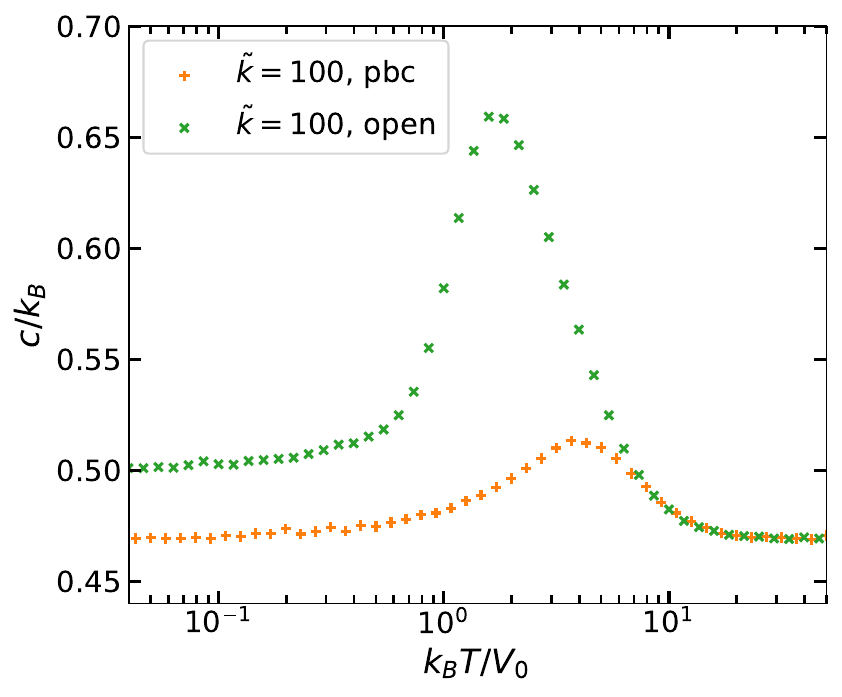}
\caption{\label{fig:cp_k100}
Configurational
specific heat $c_\text{c}$ per degree of freedom as a function of reduced temperature $k_BT/V_0$ for free atoms Frenkel Kontorova chains with strong springs, $k = 100$, which can be open (green crosses) or subjected to periodic boundary conditions (pbc, orange plus symbols). 
}
\end{figure}

As a consequence of the small barrier, sliding of the periodic chain is Stokes-like up to intermediate velocities of $\tilde{v} = 0.1$, at least for thermal energies above the tiny value of  $10^{-4}~V_0$. 
Outside the extremely cold temperature range, the equilibrium viscosity assumes a rather temperature-insensitive value of  $\eta_\textrm{eff} \approx 2 \eta_\infty$ for $T <T^*$, while $\eta \approx \eta_\infty$ when $T > T^*$ with a corrections that quickly vanishes with increasing $T$, see Fig.~\ref{fig:eff_visc_k10_obc}.
The behavior pertains to situations, which could be associated with structural lubricity in ultra-high vacuum, i.e., the contact between two incommensurate solids whose intrabulk interactions are less strong than those within the bulk, or, alternatively, with the use of solid lubricants.  

\begin{figure}[h]
\includegraphics[width=0.9\columnwidth]{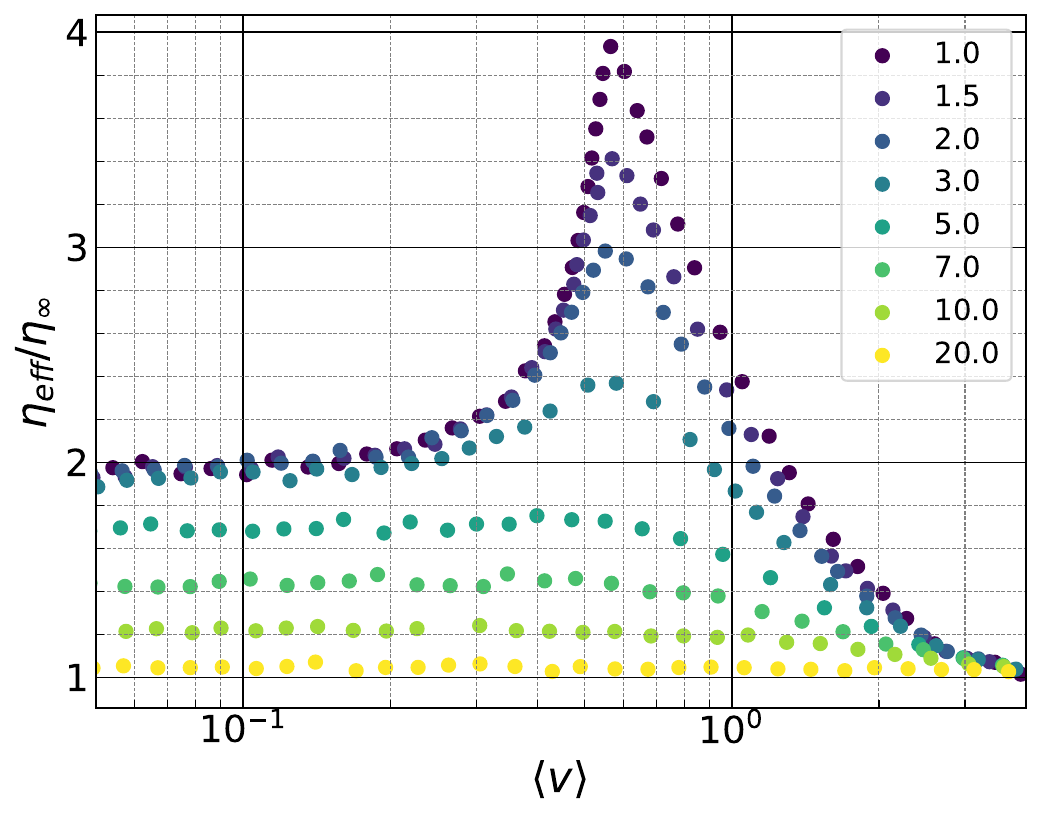}
\caption{\label{fig:eff_visc_k10_obc}
Effective viscosity of a $\tilde{k}=100$ periodic chain in a corrugated potential. 
Circles represent data obtained from simulations at constant force and reduced temperature $\tilde{T}$.
}
\end{figure}

The excess viscosity can be easily rationalized within a simple perturbation approach, which is only sketched briefly here, as more detailed treatments for this limit are well known in the literature. 
To lowest order, the external force acting on a bead can be cast as a time-dependent function under the assumption that each bead moves with its center-of-mass velocity. 
At low to intermediate velocities and $T < T^*$, the motion of the beads is  undulated only slightly due to the interactions with the substrate potential.
This leads to a small but noticeable excess of dissipation compared to smooth sliding, since increased fluctuations in velocity cause enhanced dissipation.
Each bead has sufficient time to move quasi-statically so that doubling the speed quadruples the dissipated power, which doubles the friction. 
The dissipation is enhanced, once the excitation frequency approaches a resonance frequency in the chain. 
For very high velocities, the excitation exceeds the resonance frequency and momentum as well as energy transfer into internal modes is inefficient, as in a simple harmonic oscillator. 
In this regime, the power dissipated by the chain scales as $\sim 1/{\langle v\rangle}^2$ since beads are ineratial, resulting in an effective viscosity that decays as $\eta_{\text{eff}}\sim 1/{\langle v\rangle}^4$.

When $T>T^*$, thermal fluctuations effectively smear out the substrate potential.
As a consequence, the substrate no longer modifies the motion of the beads substantially so that their mean motion is impeded predominantly by the explicitly imposed damping at very high temperature. 

\begin{figure}[htbp]
\includegraphics[width=0.9\columnwidth]{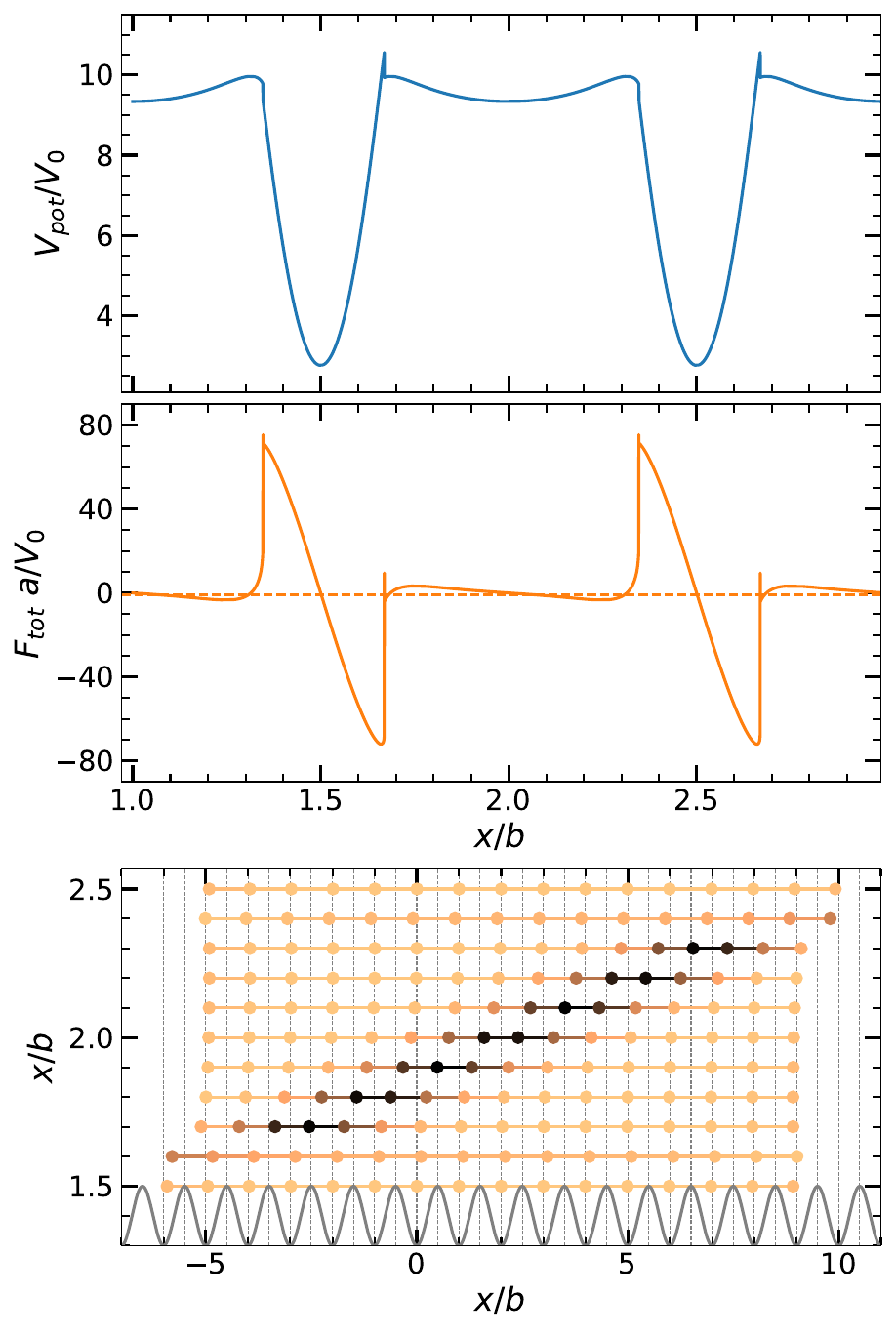}
\caption{\label{fig:barr100_obc}
Athermal dynamics of a $\tilde{k} = 100$ chain subjected to open boundary conditions. 
Potential energy $V_\text{pot}$ (top) and total lateral force $F_\text{tot}$ (middle) as a function of the moved average distance $x$
at a small sliding velocity of $v = 1e-5$.
The bottom panel shows the displacement of individual atoms on the abscissa and the average displacement on the ordinate. 
}
\end{figure}

As the final model, we consider a stiff open chain.
Its  (apparent) barrier is much larger than for the periodic chain, even noticeably larger than for uncoupled atoms, i.e., $\Delta E \gtrsim 7.79~V_0$, which is demonstrated in Fig.~\ref{fig:barr100_obc}.
This is because for the considered ratio of $b/a = 16/15$, which is near unity, a significant fraction of the chain can be close to the substrate's potential energy minimum.
However, if the spring constant were even larger than $\tilde{k} = 100$ or the $b/a$ ratio differed more clearly from unit, barriers would decrease and be dominated by end-effects. 
In fact, in the limit of $\tilde{k}\to \infty$ and $b = 1.5\,$, the barrier would disappear even for an open chain if consisted of an even number of bead and be $2\Delta V$ for an odd number of beads. 

Since the internal degrees of freedom of the open chain are rather stiff, discontinuities in the potential energy barely show under athermal, quasi-adiabatic sliding.
Thus, substantial instabilities do not occur. 
The only instabilities that is apparent to the naked eye happens for $x/b \lesssim 1.7$ when the high-energy atoms reach the leading edge of the moving chain. 
This is why the potential energy almost reflects the symmetry of the underlying lattice, i.e., it is clearly periodic in $b$ (rather than periodic in $b/P$ as for a chain subjected to periodic boundary conditions) and almost reflects the symmetry of the underlying potential.
At the same time, the potential under adiabatic center of mass advancement of a stiff but not infinitely-stiff chain is far from being single-sinusoidal in contrast to that of a single point particle moving in a single-sinusoidal field. 
A related notion has been explored in the work of Tysoe and co-authors \cite{Manzi2021TL}, who investigated how deviations from sinusoidal substrate potentials—such as piecewise parabolic forms—affect atomic-scale friction and the onset of instabilities in Prandtl–Tomlinson-type sliding models.
%

Although the instantaneous potential of the center of mass does not show a steep drop in Fig.~\ref{fig:barr100_obc}, the center of mass would pop forward once the rigid chain is pushed to the point where $V_\textrm{pop}$ starts to decrease.
Extracting dissipated energy from the energy drops is sensitive to the internal instabilities.
As a consequence, the rigid chain does not only have a high static friction force of close to $\tilde{F} = 75$, but also substantial kinetic friction.
In fact, its behavior is so similar to that of a free particle that we abstain from showing $\eta(v)$ or $\eta_0(T)$.


\section{Discussion and Conclusions}

This paper was concerned with the analysis of the thermal Frenkel-Kontorova (FK) model, with a particular, though not exclusive, focus on small sliding velocities, where thermal activation helps overcome energy barriers that hinder mass transport and intermediate velocities, where dynamics are mostly dictated by the energy landscape but not by inertia, as at extremely large velocity. 
One motivation for our study was to explore to what extent the FK model exhibits features typical of the fluid rheology of (complex) liquids or lubricants. 
We believe this endeavor was largely successful, as the following characteristics were reproduced: 
A crossover from non-Arrhenius to Arrhenius dependence of the equilibrium viscosity near a temperature where the specific heat reaches a maximum; the presence of a power-law sub-diffusive regime between the ballistic and diffusive regimes; and shear thinning, which could be described reasonably well by the Eyring equation for most studied parameters. 

While these salient features cannot be claimed with certainty to arise from the same generic mechanisms as in liquids, such a connection appears plausible. 
After all, mass transport in dense liquids also requires energy barriers to be overcome, and there is little reason to assume that the underlying physics should differ fundamentally when transport occurs via kinks in a Frenkel-Kontorova chain or through coordination defects in a disordered fluid. 
This may place us in a position to qualitatively rationalize key rheological features of liquids within the FK framework, which remains mathematically more tractable than even the simplest atomistic models, such as binary Lennard-Jones fluids. 
Naturally, the overall picture that emerges from this analysis is not fundamentally new. 
However, it may offer a more condensed and transparent formulation than previous approaches, by capturing essential phenomenology with minimal  complexity. 

The studied FK chains exhibit a peak in specific heat near the temperature below which the (mean) number of defects enabling mass transport no longer changes substantially with decreasing temperature. 
Once the number of mass-transport-inducing defects has plateaued, viscosity and diffusion display an Arrhenius-like dependence as temperature decreases. %
Moreover, while thermal activation clearly plays a primary role at the onset of shear thinning, the rate-dependent (effective) viscosity at high shear rates is primarily governed by the (local) energy drops that occur in the wake of shear-induced instabilities. 
Yet, the subsequent dependence of viscosity on shear rate or sliding velocity remains similar to that predicted by Eyring theory, which considers thermal activation alone. 
Therefore, even when Eyring theory provides a good fit, it may do so for reasons unrelated to its foundational assumptions. 

Regarding diffusion, we identify a transition from independent-particle to collective dynamics as a function of chain-stiffness. While increasing stiffness initially suppresses diffusion, a non-monotonic behavior emerges beyond a critical coupling, with diffusion approaching the uncoupled-particle limit.
    
In the tribological context, the system-dynamics is governed by fractional-instabilities till the medium coupling. However, as the chain becomes rigid, the discommensuration extends over large domain and the friction drops.

In light of the results presented in this study, the Frenkel–Kontorova model retains key features such as the Devil’s staircase and nonlinear transport behavior, reaffirming its status as a minimal yet insightful framework for capturing key rheological signatures of complex fluids—including thermally activated viscosity, shear thinning, and diffusive anomalies—within an analytically tractable and physically transparent setting.


\section*{AUTHOR DECLARATIONS}
\subsection*{Conflict of Interest}
The authors have no conflicts to disclose.

\subsection*{Data Availability}
The in-house codes and the data that support the findings of this study are available on Github \cite{Muser2025Github}. 

%




\bibliography{fk_model}

\end{document}